\documentclass{aa}

\def\OIline{[\mbox{O\,{\sc i}}]~$\lambda$6300}
\def\NIIline{[\mbox{N\,{\sc ii}}]~$\lambda$6584}
\def\SIIline{[\mbox{S\,{\sc ii}}]~$\lambda \lambda 6717,31$}
\def\OIIIline{[\mbox{O\,{\sc iii}}]~$\lambda 5007$}

\usepackage{graphicx}
\usepackage{txfonts}
\usepackage{hyperref}
\usepackage{multirow}
\usepackage{lscape}
\usepackage{amsmath}
\usepackage{placeins}

\begin{document} 

\title{The EDGE-CALIFA survey: The role of spiral arms and bars in driving central molecular gas concentrations}

   \author{Si-Yue Yu\inst{\ref{mpifr}} \thanks{Humboldt Postdoctoral Fellow}
                   \and Veselina Kalinova\inst{\ref{mpifr}}
                   \and Dario Colombo\inst{\ref{mpifr}}
                   \and Alberto D. Bolatto\inst{\ref{UM}}
                   \and Tony Wong\inst{\ref{UI}}
                   \and Rebecca C. Levy\inst{\ref{UArizona}} \thanks{NSF Astronomy and Astrophysics Postdoctoral Fellow}
           \and Vicente Villanueva\inst{\ref{UM}}
                   \and Sebastián F. Sánchez\inst{\ref{UNA}}
                   \and Luis C. Ho\inst{\ref{kiaa}}
           \and Stuart N. Vogel\inst{\ref{UM}}
           \and Peter Teuben\inst{\ref{UM}}
           \and Mónica Rubio\inst{\ref{Uchile}}
          }
    
   \institute{Max-Planck-Institut für Radioastronomie, Auf dem Hügel 69, 53121 Bonn, Germany\\ \email{syu@mpifr-bonn.mpg.de}\label{mpifr} 
   \and Department of Astronomy, University of Maryland, College Park, MD 20742, USA\label{UM}
   \and Department of Astronomy, University of Illinois, Urbana, IL 62801, USA\label{UI}
   \and Steward Observatory, University of Arizona, Tucson, AZ 85721, USA \label{UArizona}
   \and Instituto de Astronomía, Universidad Nacional Autónoma de México, A.P. 70-264, 04510 México, D.F., Mexico \label{UNA}
   \and Kavli Institute for Astronomy and Astrophysics, Peking University, Beijing 100871, China\label{kiaa}
   \and Departamento de Astronomía, Universidad de Chile, Casilla 36-D Santiago, Chile \label{Uchile}
}

\abstract{
Shocks and torques produced by non-axisymmetric structures such as spiral arms and bars may transport gas to galaxy central regions. We test this hypothesis by studying the dependence of the concentration of CO luminosity ($C_{\rm CO}$) and molecular gas ($C_{\rm mol}$) and the star formation rate ($C_{\rm SFR}$) in the central $\sim$\,2\,kpc on the strength of non-axisymmetric disk structure using a sample of 57 disk galaxies selected from the EDGE-CALIFA survey. The $C_{\rm mol}$ is calculated using a CO-to-H$_2$ conversion factor that decreases with higher metallicity and higher stellar surface density. We find that $C_{\rm mol}$ is systematically 0.22\,dex lower than $C_{\rm CO}$. We confirm that high $C_{\rm mol}$ and strong non-axisymmetric disk structure are more common in barred galaxies than in unbarred galaxies. However, we find that spiral arms also increase $C_{\rm mol}$. We show that there is a good correlation between $C_{\rm mol}$ and the strength of non-axisymmetric structure (which can be due to a bar, spiral arms, or both). This suggests that the stronger the bars and spirals, the more efficient the galaxy is at transporting cold gas to its center. Despite the small subsample size, the $C_{\rm mol}$ of the four Seyferts are not significantly reduced compared to inactive galaxies of similar disk structure, implying that the active galactic nucleus feedback in Seyferts may not notably affect the molecular gas distribution in the central $\sim$\,2\,kpc. We find that $C_{\rm SFR}$ tightly correlates with $C_{\rm mol}$ in both unbarred and barred galaxies. Likewise, elevated $C_{\rm SFR}$ is found in galaxies with strong disk structure. Our results suggest that the disk structure, either spirals or bars, can transport gas to the central regions, with higher inflow rates corresponding to stronger structure, and consequently boost central star formation. Both spirals and bars play, therefore, an essential role in the secular evolution of disk galaxies.
}

\maketitle

\section{Introduction}

Molecular gas is a key material for star formation. Observations of molecular gas provide an important tool for exploring how structures or processes influence galaxy evolution. Molecular gas is typically traced through CO observations \citep[e.g.,][]{Sakamoto1999a, Sheth2005, Kennicutt2007, Bigiel2008, Leroy2009, Wilson2012, Bolatto2017, Sorai2019, PHANGS2021}. Spatially resolved CO data furnish essential information about the global gaseous properties of galaxies, such as the radial distribution of molecular gas.

Spiral arms and bars are the most common features in disk galaxies. In the local universe, nearly 60\% of disk galaxies have a bar \citep[e.g.,][]{Aguerri2009, Simon2016}.  Models and simulations show that the non-axisymmetric bar gravitational potential drives gas flow toward the galaxy central region along the bar dust lane \citep[e.g.,][]{Athanassoula1992a, Athanassoula1992b, Regan1999, Sheth2000, Sheth2002, Regan2004, KWT2012, Combes2014, Fragkoudi2016, Tress2020}.  In a galaxy with a well-defined inner Lindblad resonance (ILR), the gas transported by the bar accumulates at the ILR at a distance of a few kiloparsecs from the center \citep{Athanassoula1984, Jenkins1994}; later, the response of gas to forcing by the bar can give rise to a nuclear spiral or bar pattern in the gas, which can transport the gas to the proximity of the central black hole \citep{Englmaier2000, Englmaier2004}. Observations have confirmed bar-driven gas transport \citep[e.g.,][]{Mundell1999, Combes2003, Zurita2004, Fathi2006, Jogee2006, Haan2009, Querejeta2016}. The rates of gas inflow range from 0.01 to 50\,$M_{\odot}$/yr \citep{Regan1997, Haan2009, Querejeta2016}. If the gas accumulation in the center is faster than the gas consumption, the inflow of gas will result in a centrally concentrated distribution of gas. Based on 20 galaxies from the Nobeyama Radio Observatory–Owens Valley Radio Observatory (NRO-OVRO) survey \citep{Sakamoto1999a},  \cite{Sakamoto1999b} found that the molecular gas concentrations in the central 1 kpc are systematically higher in barred spirals than in unbarred spirals. Consistently, with a larger and more diverse sample of 44 galaxies, the Berkeley-Illinois-Maryland Association (BIMA) CO Survey of Nearby Galaxies (SONG) confirmed that high central gas concentrations are more common in barred galaxies \citep{Sheth2005}. Further, the central gas concentrations are higher in galaxies hosting a stronger bar \citep{Kuno2007}, suggesting that stronger bars drive more gas inflow \citep{Regan2004, Hopkins2011, KWT2012}. \cite{Komugi2008} showed that early-type galaxies harbor larger central concentrations, a trend they attribute to the effect of bulges.

The gas inflow driven by the bar leads to an increase in gas density at the center, which enhances the central star formation activity \citep{Ho1997, Martinet1997, Aguerri1999, Sheth2005, Regan2006, Schinnerer2006, Wang2012, Combes2014, Zhou2015, Lin2017, Chown2019, Lin2020, Wang2020, Tress2020, Sormani2020}. Stronger bars tend to have higher levels of central enhancements of star formation rates \citep[SFRs;][]{Zhou2015, Lin2017}. Nevertheless, some strongly barred galaxies could have normal or suppressed central star formation \citep{Martinet1997, Wang2012, Wang2020, Consolandi2017, Simon2020}. In these cases, the bar may have transported gas to the center, where the gas has been consumed by star formation.

Since high central gas concentrations are more common in barred galaxies, the gas inflow driven by spiral arms should typically be weaker compared to the bar effect. A signature of spiral-driven gas inflow has been detected. \cite{Regan2006} showed that there are two unbarred galaxies  among the known six disk galaxies that have central excess in the 8\,$\mu$m and CO emission above the inward extrapolation of an exponential disk; hence, their molecular gas content is centrally concentrated. In \cite{Sheth2005}, a few unbarred spirals show relatively high gas concentrations, although none of them reach the very high values seen in some barred galaxies. Recently, \cite{Yu2022} analyzed central star formation in 2779 nearby unbarred star-forming (SF) disk galaxies; they find higher central SFRs in galaxies with strong spiral arms, implying that strong spiral arms may transport gas to the center. The strength of spiral arms, analogous to the role of bar strength, may be a key factor in studying the impact of spiral arms. This point of view is supported by models and simulations, which have shown that the non-axisymmetric spiral potential provides an efficient mechanism for transferring angular momentum, causing a radial inflow of gas. Firstly, the quasi-static density waves \citep{LinShu64} predict that spiral potential generated by old stars induces large-scale galactic shocks on a gas cloud as the gas cloud crosses the arm \citep{Roberts1969}. The large-scale galactic shocks dissipate angular momentum, causing the gas cloud in orbital motions to move radially inward inside the corotation resonance \citep{Kalnajs1972, Roberts1972, Lubow1986, Kim2014, Baba2016, Kim2020}. Secondly, gravitational torque of the non-axisymmetric spiral potential generated by old stars drives gas inflow \citep{Kim2014, Baba2016, Kim2020}. Thirdly, gravitational torque of the gaseous component has an additional minor contribution to gas inflow \citep{Kim2014}. \cite{Kim2014} find that the rate of gas inflow to the central region, caused by a combination of the above three processes driven by spiral arms, is higher in galaxy models with stronger and more slowly rotating arms.

The dependence of the gas inflow rate on the spiral arms strength suggests that the stronger the spiral arms, the higher the molecular gas concentration. If this is true, and given the known correlation between bar strength and gas concentration \citep{Kuno2007}, the impact of spirals and bars may be uniformly described by the strength of the non-axisymmetric disk structure. The arms and bars jointly influence the radial distribution of molecular gas. In barred galaxies, the arms first transport gas to a radial extent within the bar and the bar successively drives the gas toward the center.

A constant CO-to-H$_2$ conversion factor ($\alpha_{\rm CO}$) has usually been adopted to convert CO integrated intensities to molecular gas mass surface densities and then derive molecular gas concentrations \citep{Sakamoto1999b, Sheth2005, Kuno2007}. However, studies have shown that the $\alpha_{\rm CO}$ depends on the physical properties of the environment where the gas clouds are embedded \cite[e.g.,][]{Narayanan2012, Bolatto2013}. Hydrodynamical simulations suggest that the $\alpha_{\rm CO}$ could vary by orders of magnitude in different environments \citep{Feldmann2012, Gong2020}. The $\alpha_{\rm CO}$ tends to be lower if molecular clouds have decreased density and/or increased temperature \citep[e.g.,][]{Bolatto2013}, have increased velocity dispersion \citep[e.g.,][]{Watanabe2011}, or have increased metallicity \citep[e.g.,][]{Israel1997, Wolfire2010}. Galaxy centers tend to have lower $\alpha_{\rm CO}$, perhaps due to higher temperatures, higher metallicity, and/or dynamical effects in the centers \citep[e.g.,][]{Strong2004, Sandstrom2013, Sanchez2014, Israel2020, Teng2022}. A central $\alpha_{\rm CO}$ depression reduces the contribution of molecular gas in the central region when deriving molecular gas concentrations, and the true molecular gas concentrations are therefore lower than those derived in the literature. The degree of central depression in $\alpha_{\rm CO}$ varies from galaxy to galaxy, exerting different levels of influence on the molecular gas concentrations. It is not known how a variable $\alpha_{\rm CO}$ quantitatively influences the molecular gas concentrations and the related results.

In this work we aim to understand the influence of variable $\alpha_{\rm CO}$ on molecular gas concentrations and study the impact of disk (spiral+bar) structure on the radial distribution of molecular gas and star formation. The paper is structured as follows. Section~\ref{data} describes the data. Section~\ref{Methods} presents the methods for measuring the central concentrations of CO luminosity, molecular gas and the star formation rate, and for calculating the strength of non-axisymmetric disk structure. Sections~\ref{results} and \ref{discussion} show the results and discussions, respectively. A summary of the main conclusions is given in Sect.~\ref{conclusions}.

\section{Observational material}\label{data}
\subsection{Sample and data}

The Extragalactic Database for Galaxy Evolution survey \citep[EDGE;][]{Bolatto2017} observes $^{12}$CO $J$=1\textendash0 and $^{13}$CO $J$=1\textendash0 in 126 nearby galaxies using a combination of D and E configurations (D+E) of the Combined Array for Research in Millimeter-wave Astronomy (CARMA) interferometer. It has a typical synthesized beam of $\sim$\,4$\farcs$5, corresponding to $\sim$\,1.5\,kpc for the EDGE sample.  EDGE galaxies are selected from the Calar Alto Legacy Integral Field Area (CALIFA) surveys \citep[][]{Sanchez2012}, and these galaxies form a representative sample of the SF galaxies in CALIFA, although most of them were selected for their high mid-infrared flux and convenience of scheduling observations \citep{Bolatto2017}. The average RMS noise in EDGE data cubes gives $\sim$\,50\,mK, corresponding to a typical 3\,$\sigma$ sensitivity of molecular gas mass surface density: $\sim$\,11\,$M_{\odot}$\,pc$^{-2}$ \citep{Bolatto2017}. We used the $^{12}$CO $J$=1\textendash0 moment 0 maps from D+E data cubes to quantify the central concentrations of CO luminosity and molecular gas probed in the present paper. The significant noise in the maps was rejected by applying a blanking mask through the smooth masking approach \citep{Bolatto2017}. We used smoothed masks instead of dilated masks, as the smoothed-mask moment maps provide a more complete accounting of the total CO flux \citep{Bolatto2017}. The maps were resampled so that each resolution element has $\sim$\,4 pixels \citep{Utomo2017}.

 The integral field unit (IFU) data from the third CALIFA data release was used to characterize properties of metallicity, stars, and star formation. We used maps of emission line of H$\alpha$, H$\beta$, \OIline, \NIIline, \OIIIline, and \SIIline, and maps of stellar surface density ($\Sigma_{\star}$), obtained from the {\tt PIPE3D} pipeline \citep{Sanchez2016}. The oxygen abundance (O$/$H) derived based on O3N2 ratio is acquired from \cite{Marino2013}. The maps were re-gridded and smoothed to match the pixel scale and resolution of the EDGE data \citep{Utomo2017}. Pixels with signal-to-noise ratio (S$/$N) $<$\,2 and those occupied by foreground stars were set to blank.

From the parent EDGE sample, we excluded galaxies that were not suitable for studying disk structures based on visual inspection. First, we excluded elliptical galaxies, where a large-scale disk is not present. Second, we rejected galaxies whose optical image is severely contaminated by bright foreground stars or nearby galaxies. Galaxies presenting tidal tails, which are thus interacting galaxies, were rejected. This process was to ensure the robustness of quantifying the non-axisymmetric disk structure, that is, spiral arms and bars. Finally, nearly edge-on galaxies were removed. This selection process was done by visual inspection, rather than using a cut in inclination angle derived from their axis ratio, because the axis ratio would be underestimated for early-type edge-on galaxies with spherical stellar haloes. Disk structures are invisible and unrecoverable in edge-on galaxies. Furthermore, three galaxies without CO detection in their central regions ($R$\,$<$\,0.2\,$R_e$, where $R$ is the galactocentric radius in the face-on viewing angle and $R_e$ is the optical half-light radius) were removed because their gas concentrations are not available (Sect.~\ref{Ccomol}). Our final sample consists of 57 galaxies (Table~\ref{tdata}).

\begin{figure}
        \centering
        \includegraphics[width=0.48\textwidth]{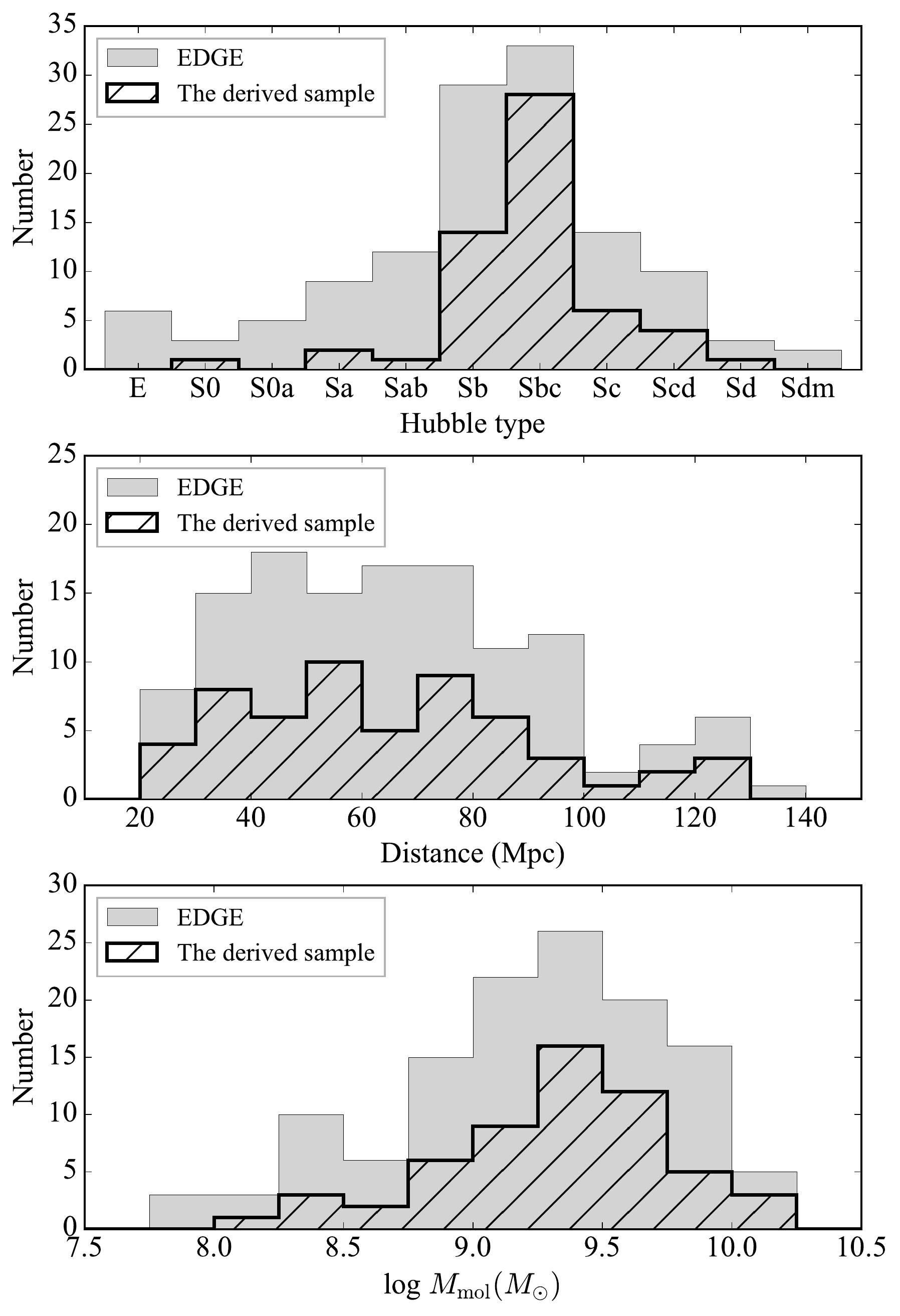}
        \caption{Histogram of Hubble types, distance, and total molecular gas mass ($M_{\rm mol}$) of galaxies in the EDGE parent sample and our derived sample. 
        }
        \label{Hubble}
\end{figure}

Figure~\ref{Hubble} compares Hubble types, distances, and global molecular gas masses \citep{Bolatto2017} of our derived sample and the EDGE parent sample. Our sample favors late-type (Sb--Sd) galaxies, showing a similar Hubble-type distribution to late-type galaxies in the EDGE parent sample. We include one S0 galaxy, NGC~5784. S0s are disk galaxies without spiral arms by definition, but some S0s could have faint arms \citep{Kormendy2012, YuHo2020}. NGC~5784 has molecular gas mass of $10^{9.4}$\,$M_{\odot}$ \citep{Bolatto2017}. Its gas content was not removed, contrary to the scenario that progenitors of S0s should lose gas and then structurally subside \citep{vandenBergh1976}. The vanished arm structure in NGC~5784 therefore represents weak effect of spiral arms on radial distribution of gas. This galaxy is included in the sample. The histograms of distance and total molecular gas mass are close to those of the EDGE parent sample, indicating that there is no significant bias in these two parameters for our sample. All galaxies have available 2D multicomponent photometric decomposition from \cite{JMA2017}. We use the bulge half-light radius ($R_{\rm bul}$) measured at {\it r}-band from \cite{JMA2017} when determining disk-dominated regions in Sect.~\ref{Methods}. Other parameters are derived in Sect.~\ref{Methods}.  We use the bar identification from \cite{Walcher2014}. 34 (60\%) objects are barred galaxies (SAB or SB), and 23 (40\%) objects are unbarred galaxies (SB).

\subsection{Nuclear activity}

The proposed relationship between central concentrations of molecular gas and strengths of non-axisymmetric disk structure may provide a tool to investigate effects of active galactic nucleus (AGN) feedback on the radial distribution of molecular gas. Measurement of the SFR based on H$\alpha$ emission suffers from contamination by emission from the AGN, and AGN hosts should therefore not be involved when probing galaxy central star formation. For the above two reasons, we classify the dominant energy source for the galaxies in our sample.

We use the Baldwin-Phillips-Terlevich (BPT) diagram \citep{BPT1981} to search for AGN candidates. The BPT diagram makes use of the emission line ratios ${\rm [OIII]/H\beta}$, ${\rm [NII]/H\alpha}$, ${\rm [SII]/H\alpha}$, and ${\rm [OI]/H\alpha}$. We classify galaxies as pure SF galaxies if the central ($R$\,$<$\,0.2\,$R_e$) pixels locate below the pure star formation line of \cite{Kauffmann2003AGN} on the ${\rm [NII]/H\alpha}$ versus ${\rm [OIII]/H\beta}$ diagram. The reason for using 0.2\,$R_e$ as inner boundary is to be consistent with the definition of central surface density of molecular gas mass and SFR (see Sect.~\ref{Methods}). Composite line ratios are caused by a combination of star formation and AGN activity \citep{Kewley2006, Yuan2010}. We classify galaxies as composite galaxies if the central pixels lie between the pure star formation line of \cite{Kauffmann2003AGN} and the extreme starburst classification line of \cite{Kewley2001}. 

The remaining galaxies are further classified as candidate Seyferts, low-ionization nuclear emission line regions (LINERs), and ambiguous galaxies, using the Seyfert-LINER demarcation lines on the ${\rm [SII]/H\alpha}$ versus ${\rm [OIII]/H\beta}$ and ${\rm [OI]/H\alpha}$ versus ${\rm [OIII]/H\beta}$ diagram \citep{Kewley2006}. Candidate Seyferts and LINERs are those galaxies whose central pixels lie above and below the two demarcation lines, respectively. This procedure results in 6 candidate Seyferts. No LINERs are found, perhaps due to the EDGE sample selection requiring the galaxies to be mid-infrared bright. If galaxies do not meet the above criteria, we classify them as ambiguous galaxies. The BPT diagram has been widely adopted to identify AGNs \citep[e.g.,][]{Kewley2006, Ho2008, Koss2010}. However, diffuse ionization associated with old-stellar populations \citep{Binette1994, Singh2013, Gomes2016} may also populate the regions for AGNs in BPT diagram. \cite{Lacerda2018} find that regions where the ionization is dominated by hot low-mass evolved stars have H$\alpha$ equivalent widths  (EW$_{\rm H\alpha}$) $<$\,3\AA. It is therefore possible to distinguish between presence of an AGN from old-stellar ionization by introducing a cut in the EW$_{\rm H\alpha}$ \citep{Lacerda2018, Lacerda2020, Levy2019, Kalinova2021}. Following the strategy in \cite{Lacerda2020}, we require that AGNs have central EW$_{\rm H\alpha}$ larger than 3\,\AA. We then classify the candidate AGNs with central EW$_{\rm H\alpha}$\,$>$\,3\,\AA\ as real AGNs. Candidates with central EW$_{\rm H\alpha}$\,$<$\,3\,\AA\ could have ionization contaminated by old-stellar populations and are classified as ambiguous galaxies.

The above selection results to 30 pure SF galaxies, 15 composite galaxies, 4 Seyferts (NGC~2410, NGC~2639, NGC~6394, UGC~3973), and 8 ambiguous galaxies. Our AGN classification is consistent with the classification in \cite{Lacerda2020}. Specifically, UGC~3973 is a type~1 AGN, and NGC~2410, NGC~2639, and NGC~6394 are type~2 AGNs \citep{Lacerda2020}.

\begin{figure*}
        \centering
        \includegraphics[width=0.8\textwidth]{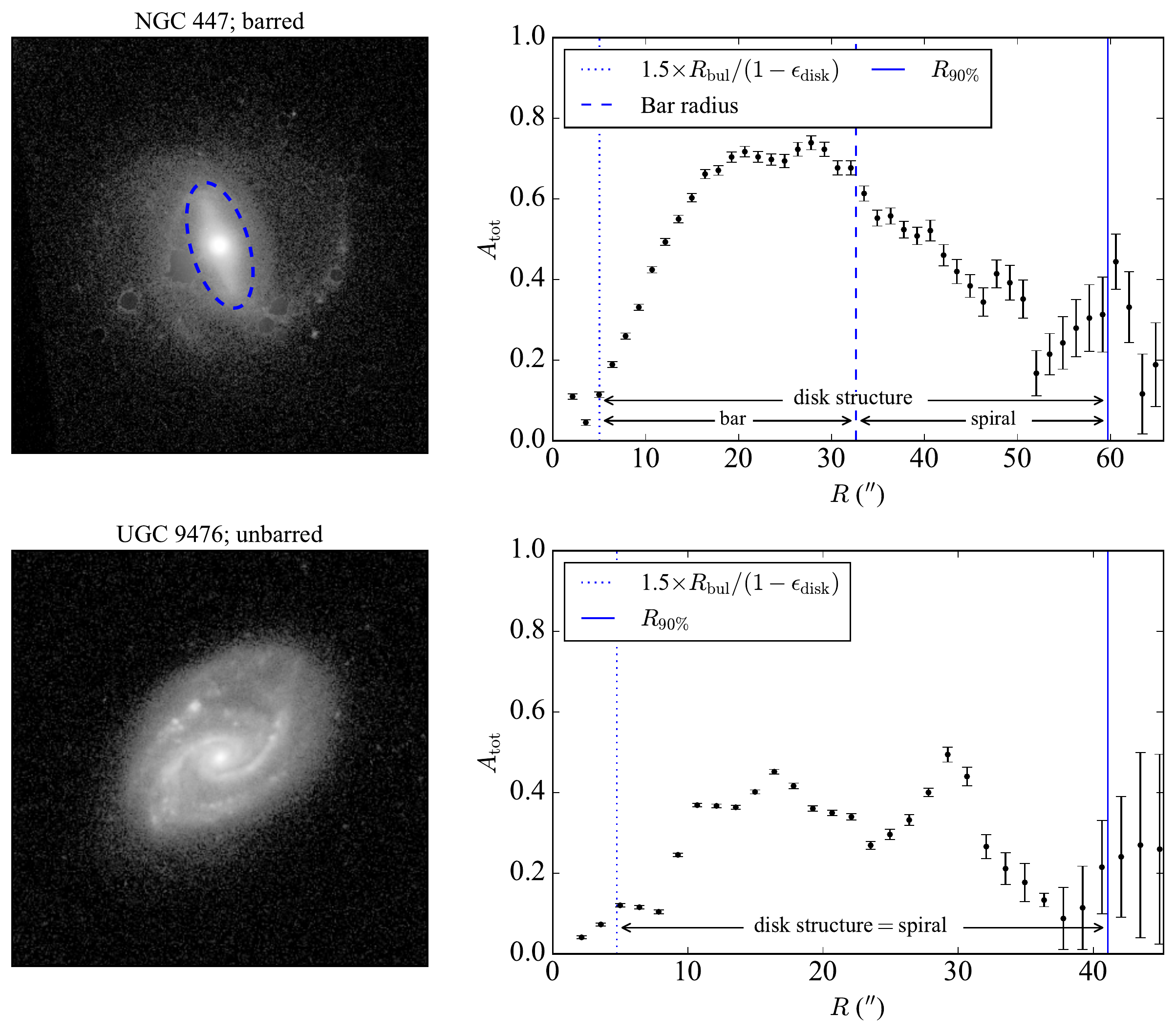}
        \caption{Illustration of how to measure the strength of non-axisymmetric disk structures. The left two panels show the star-cleaned {\it r}-band image of NGC~447 and UGC~9476. The dashed blue ellipse in the image of NGC~447 denotes its bar (Appendix~\ref{ISO}). The right panels show radial profiles of $A_{\rm tot}$. The galactocentric radius at which the bulge terminates and the disk starts to dominate ($1.5\times R_{\rm bul}/(1-\epsilon_{\rm disk})$, where $\epsilon_{\rm disk}$ is the disk-averaged ellipticity) is indicated by the dotted line. The bar radius is indicated by the dashed line for NGC~447. The $R_{90\%}$, enclosing the majority of structure, is indicated by the solid line. We compute $s_{\rm disk}$ as the mean $A_{\rm tot}$ over the disk-dominated region between $1.5\times R_{\rm bul}/(1-\epsilon_{\rm disk})$ and $R_{90\%}$, and define the strength of non-axisymmetric disk structure as $2+\log s_{\rm disk}$. The disk-dominated region for barred galaxies contains spirals and bars, but that for unbarred galaxies contains only spirals. 
        }
        \label{exmp}
\end{figure*}
%--------------------------------------------------------------------

\section{Methods}\label{Methods}
In this section we first describe how to measure the strength of non-axisymmetric disk structure, which traces average strength of spiral arms for unbarred galaxies and average strength of spirals and bars for barred galaxies. Later, we present the procedure to quantify central concentrations of CO luminosity, molecular gas, and SFR.

\subsection{Strength of non-axisymmetric disk structures}\label{sds}
We analyze the disk (spiral+bar) structure using the {\it r}-band images from the Data Release 12  \citep{Alam2015} of the Sloan Digital Sky Survey \citep[SDSS;][]{York2000}. To reduce the image data, we use {\tt SEP}\footnote{https://sep.readthedocs.io/en/v1.1.x/} \citep{Bertin1996, Barbary2016} to automatically generate a preliminary mask of foreground stars, and then manually mask out the stars, mainly inside the galaxy, that {\tt SEP} has missed.  After applying the mask to reject contamination, we average the fluxes over the region where the flux profile flattened to calculate background and then subtract this value from the image. 

For each galaxy we run {\tt IRAF} task {\tt ellipse} with an exponential step of 0.05 to obtain isophotal parameters such as isophotal ellipticity ($\epsilon$) and position angle (PA). We estimate the intrinsic galaxy light affected by the masked region using the task {\tt bmodel}, which builds a smooth representation of the galaxy light based on the isophotes, and then use these values to fill in the masked region to produce the star-cleaned images. The top-left and bottom-left panel of Fig.~\ref{exmp} illustrates the star-cleaned images of NGC~447 and UGC~9476, respectively. The galaxy disk-averaged ellipticity $\epsilon_{\rm disk}$ and position angle PA$_{\rm disk}$ are obtained by averaging the profiles of $\epsilon$ and PA over the outer disk or by minimizing the real part of the $m$\,$=$\,$2$ mode of 2D Fourier spectra at a radial wavenumber of zero (for details, see \citealt{Grosbol2004} and \citealt{Yu2018}). The measured $\epsilon_{\rm disk}$ and PA$_{\rm disk}$ are listed in Table~\ref{tdata}. Our measured $\epsilon_{\rm disk}$ and PA$_{\rm disk}$ are consistent with those from HyperLEDA \citep{Makarov2014} with a standard deviation of 0.07 and 10$\degr$, respectively. The 0.07 and 10$\degr$ are taken as the measurement uncertainty of $\epsilon_{\rm disk}$ and PA$_{\rm disk}$, respectively. We then apply elliptical apertures with the $\epsilon_{\rm disk}$ and PA$_{\rm disk}$ to derive $R_{50\%}$ (denoted as $R_e$ hereafter) and $R_{90\%}$, which contains 20\% and  90\% of total flux of the galaxy.

In order to quantify the non-axisymmetric disk (spiral+bar) structure, we first define a disk-dominated region for each galaxy.  A galaxy is bulgeless if a bulge component is not included in the 2D multicomponent photometric decomposition in \cite{JMA2017}. The inner boundary of the disk-dominated region is set to $1\farcs4/(1-\epsilon_{\rm disk})$ for bulgeless galaxies. The $1\farcs4$ is the typical SDSS {\it r}-band point spread function (PSF) full width at half medium. For galaxies hosting a bulge, the inner boundary is set to $1.5\times R_{\rm bul}/(1-\epsilon_{\rm disk})$. The factor of $(1-\epsilon_{\rm disk})$ is to take into account the geometric difference between the round bulge or PSF and the inclined disk. The factor of 1.5 is to ensure that the majority of bulge flux is excluded. The outer boundary of the disk-dominated region is set to $R_{90\%}$, which by definition encloses the majority of the disk \citep{Yu2018}.

 The strength is the most fundamental property of spirals and bars, and therefore of the disk structure. We adopt a homogeneous method to quantify the disk (spiral+bar) structure based on 1D Fourier decomposition of azimuthal intensity distribution along each isophote \citep{Elmegreen1989, Laurikainen2004, Elmegreen2011, Rix1995, Grosbol2004, Durbala2009, Baba2015, Kendall2011, Kendall2015, Yu2018, YuHo2020}. We run the {\tt ellipse} task with ellipticity fixed to $\epsilon_{\rm disk}$, position angle fixed to PA$_{\rm disk}$, and with a linear step of 1$\farcs$4. The resulting azimuthal intensity distribution, $I(R,\theta)$, along an isophote at radius $R$ is decomposed through
\begin{eqnarray}
        I(R,\, \theta) = I_{0}(R) + \sum_{m=1}^{6} I_{m}(R)\, \cos [m(\theta + \phi_{m})],
\end{eqnarray}
\noindent
where $I_{0}$ is the azimuthally averaged intensity and represents the axisymmetric disk, and $I_m$ is the Fourier amplitude, which measures the structure amplitude. The relative Fourier amplitude is defined as 
\begin{eqnarray}
        A_m(R) = \frac{I_m(R)}{I_0}.
\end{eqnarray}
\noindent
 The $m$\,$=$\,$2$ is often used to quantify the strength of bars or spirals \cite[e.g.,][]{Grosbol2004, Elmegreen2007, Durbala2009}. However, since the shape of $I(R,\, \theta)$ is not an exact cosine, a bar or spiral would contribute to higher order modes \citep{Rix1995}. Following the strategy in \cite{Yu2018}, we define relative amplitude of structure as the quadratic sum of the relative Fourier amplitude of $m$\,$=$\,2, 3, and 4 modes:
\begin{equation}
  A_{\rm tot}=\sqrt{A_2^2+A_3^2+A_4^2}.
\end{equation}
\noindent
The top-right and bottom-right panels of Fig.~\ref{exmp} illustrate the $A_{\rm tot}(R)$ for NGC~447 and UGC~9476, respectively. The disk-dominated regions of NGC~447 and UGC~9476 are marked by the region between the vertical dotted (the end of bars or bulges) and the vertical solid (the end disks) lines in Fig.~\ref{exmp}.  The bar of NGC~447, quantified using isophotal analysis described in Appendix~\ref{ISO}, is illustrated by the dashed ellipse in the top-left panel and the bar radius is shown as the vertical dashed line in the top-right panel. The disk-dominated region for barred galaxies contains spirals and bars, but that for unbarred galaxies contains only spirals. The average relative amplitude of the disk structure, $s_{\rm disk}$, is defined as the mean $A_{\rm tot}$ over the disk dominated region:
\begin{equation}
  s_{\rm disk} = {\rm avg}[A_{\rm tot}(R_d)],
\end{equation}
\noindent
where $R_d$ belongs to the disk-dominated region. $s_{\rm disk}$\,$=$\,0.49 for NGC~447 and $s_{\rm disk}$\,$=$\,0.29 for UGC~9476. The strength of non-axisymmetric disk structure is then defined as $2+\log\,s_{\rm disk}$.  The logarithmic format is used as the spiral or bar effect may be highly nonlinear \citep{Yu2021}. The value of 2 is added to make the quantity greater than zero. Uncertainty of the structure strength is estimated by considering the standard error of the mean value, error of $\epsilon_{\rm disk}$, and error of PA$_{\rm disk}$. The derived  $2+\log\,s_{\rm disk}$ and uncertainties are listed in Table~\ref{tdata}. In strongly barred galaxies, the disk structure strength is driven by bars, because the strong bar is long and dominates the disk-dominated region. However, in weakly barred galaxies, the disk structure strength is driven by spirals, because the weak bar is short \citep{Elmegreen2007} and the spiral arms rule the disk-dominated region. Despite the different radial extent occupied by bars and spirals, the disk may facilitate the formation of arms and bars to a similar level of strength (\citealt{Salo2010, Simon2019}, but see \citealt{Athanassoula2009}). The disk structure strength ($2+\log\,s_{\rm disk}$) traces average spiral strength for unbarred galaxies, and average spiral and bar strength for barred galaxies.

The bar strength has been quantified in the literature by its ellipticity or length \citep{Martin1995, Martinet1997, Aguerri1999, Marinova2007, Barazza2008, Barazza2009, Gadotti2009, Li2011}. It is instructive to compare bar strengths defined in our framework with these classic indicators. We calculate average relative Fourier amplitude over the region occupied by the bar: $s_{\rm bar}$ and then compute the bar strength: $2+\log s_{\rm bar}$. We find a good relation where the value of $2+\log s_{\rm bar}$ gets higher with increasing bar ellipticity (Fig.~\ref{app_ebar}) and bar length (Fig.~\ref{app_Lbar}). A more sophisticated measure of bar strength is to estimate the gravitational torque \citep{Block2004, Laurikainen2004, Buta2005}. Nevertheless, it has been shown that bar torques are very well correlated with bar ellipticity \citep{Laurikainen2002, Block2004}. We therefore argue that our approach for quantifying the bars is validated.

\subsection{Central concentrations of CO luminosity and molecular gas}\label{Ccomol}

We use the EDGE $^{12}$CO $J$=1\textendash0 integrated intensity to derive the central concentrations of CO luminosity and molecular gas. Pixels without CO detections or with a S$/$N of less than 1 are blanked. Using S$/$N of 3 as the threshold yields consistent gas concentrations (Pearson correlation coefficient $\rho>0.9$). \cite{Sakamoto1999b} and \cite{Sheth2005} defined the molecular gas concentration as the ratio of the central 1\,kpc molecular gas surface density to the density average within the diameter at 25\,mag$/$arcsec$^2$ ($D_{25}$; $R_{25}=D_{25}/2$). \cite{Kuno2007} suggested using a fraction of the galaxy diameter as the inner aperture, instead of a fixed radius, to avoid a possible dependence on the galaxy size. Larger galaxies tend to have longer bars \citep{Gadotti2009}, which may have a more prominent effect on the gas distribution \citep{Kuno2007}. \cite{Chown2019} used a different definition, the ratio of optical half-light radius to CO flux half-light radius. This definition measures gas concentrations in central $\sim$\,8\,kpc for our sample. However, an aperture of this size is too large and would cancel out the difference in central gas concentration between barred and unbarred galaxies \citep{Kuno2007}, making this definition undesirable.

 The molecular gas concentrations calculated by these authors are based on a constant $\alpha_{\rm CO}$, and therefore have bias and a certain amount of uncertainty if $\alpha_{\rm CO}$ varies from location to location within the galaxy. To be precise, they measured central concentrations of CO luminosity. Using the projection parameters, $\epsilon_{\rm disk}$ and PA$_{\rm disk}$, to take into account the inclination effect and, similar to the strategy in \cite{Kuno2007}, we define the central concentration of CO luminosity ($C_{\rm CO}$) as the ratio of the average central CO intensity density ($I_{\rm CO}(R\leq 0.2\,R_e)$) to the disk-averaged value ($I_{\rm CO}(R\leq R_{e})$): 
\begin{equation}\label{Cco}
  C_{\rm CO}=\log \frac{ I_{\rm CO}(R\leq 0.2\,R_e)  }{ I_{\rm CO}(R\leq R_e)}.
\end{equation}
\noindent
The galaxy size is characterized by the $R_e$. The inner radius of 0.2\,$R_e$ is chosen to ensure that the aperture for each galaxy is larger than the beam size of 4$\farcs$5. The average 0.2\,$R_e$ of galaxies in our sample corresponds to 1.1\,kpc (2.2\,kpc in diameter), consistent with the inner radius used in \cite{Kuno2007}. Equation~(\ref{Cco}) therefore measures CO luminosity concentration in the central $\sim$\,2\,kpc. The $C_{\rm CO}$ should be slightly smaller than the concentration computed using 1\,kpc as the inner aperture diameter, especially for highly concentrated CO distributions. We refrain from doing the same calculation as in \cite{Sakamoto1999b} and \cite{Sheth2005}, because the central 1\,kpc is smaller than the beam size for most of the galaxies in our sample.

\begin{figure}
        \centering
        \includegraphics[width=0.45\textwidth]{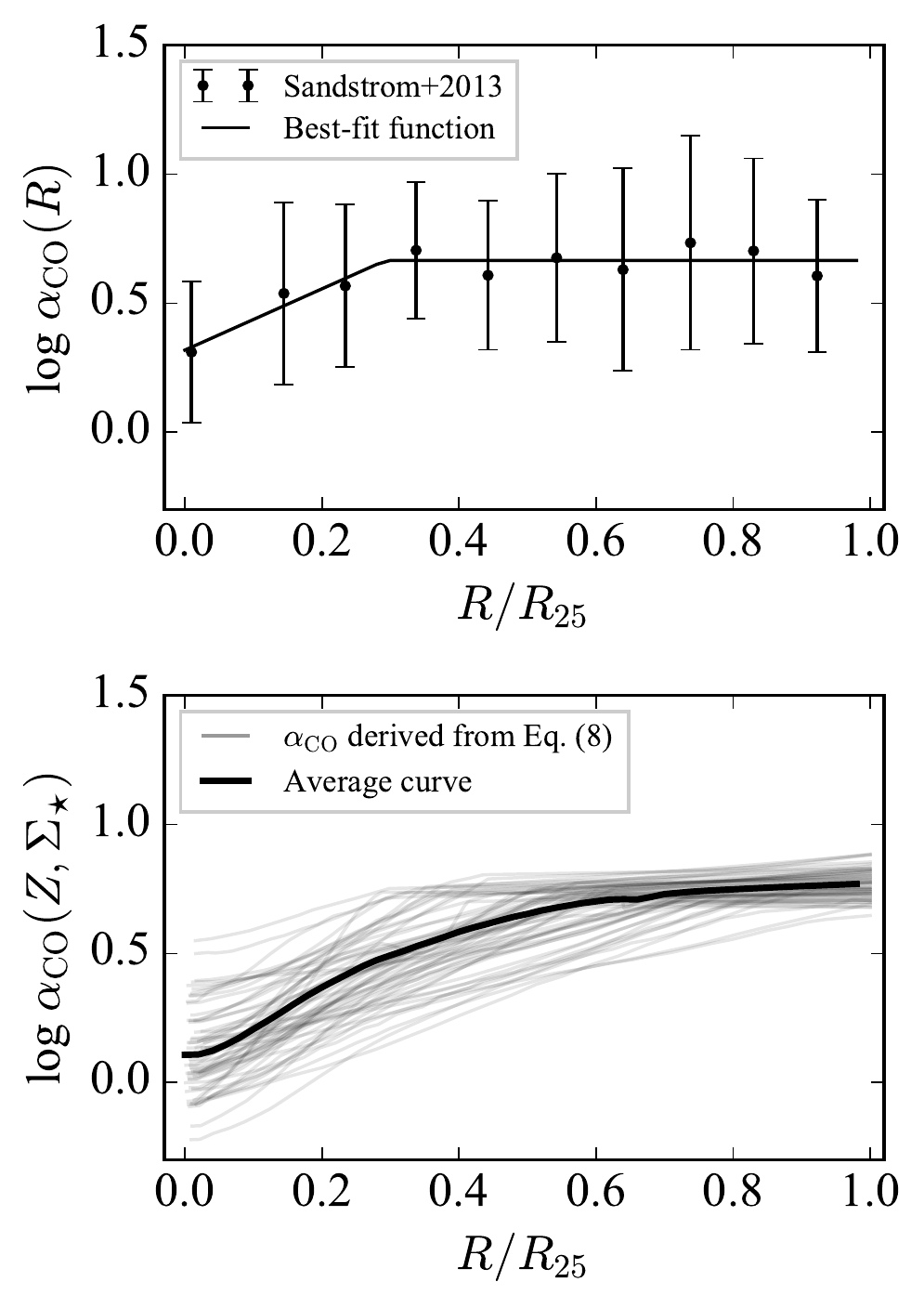}
        \caption{$\alpha_{\rm CO}$ plotted against galactocentric radius normalized by $R_{25}$. {\it Top panel:}  $\alpha_{\rm CO}(R)$  obtained by multiplying the empirical normalized average profile of $\alpha_{\rm CO}$ adopted from \cite{Sandstrom2013} by Milky Way $\alpha_{\rm CO, MW}=4.4$\,$M_{\odot}$\,pc$^{-2}$\,(K\,km\,s$^{-1}$)$^{-1}$; the errors are also from \cite{Sandstrom2013} and the black curve is the best-fit function. {\it Bottom panel:}  $\alpha_{\rm CO}(Z,\Sigma_{\star})$, marked in gray, derived from Eq.~(\ref{aco_zsig}), which considers the decrease in $\alpha_{\rm CO}$ driven by higher metallicity ($Z$) and stellar surface density ($\Sigma_{\star}$); the black curve denotes the mean $\alpha_{\rm CO}$ for a given radius. 
        }
        \label{XCO}
\end{figure}

The integrated intensity was converted to molecular surface density using a variable, $\alpha_{\rm CO}$. We considered two functional forms of $\alpha_{\rm CO}$ (i.e., $\alpha_{\rm CO}(R)$ and $\alpha_{\rm CO}(Z,\Sigma_{\star})$) to probe their effect on the calculation of molecular gas concentrations. \cite{Sandstrom2013} solved for $\alpha_{\rm CO}$ by assuming that dust and gas are well mixed and the dust-to-gas ratio is approximately constant within a given region of $\sim$\,1\,kpc. They find that the $\alpha_{\rm CO}$ in the center is on average a factor of 2 lower than the rest of the galaxy. We adopted the empirical normalized average radial profile of $\alpha_{\rm CO}$ from \cite{Sandstrom2013} and multiplied the profile by the Milky Way $\alpha_{\rm CO, MW}=4.4$\,$M_{\odot}$\,pc$^{-2}$\,(K\,km\,s$^{-1}$)$^{-1}$ to obtain the un-normalized profile $\alpha_{\rm CO}(R)$ (Fig.~\ref{XCO}). We fit the data using a broken function:
\begin{equation}\label{aco_r}
  \log \alpha_{\rm CO}(R) =
    \begin{cases}
      k\times(R/R_{25}) + b & \text{if $R/R_{25}\leq R_c$,}\\
      k\times(R_c/R_{25}) + b & \text{if $R/R_{25}> R_c$.}\\
    \end{cases}       
\end{equation}
The best-fit parameters are: $k$\,$=$\,$1.2$, $b$\,$=$\,$0.3$, and $R_c$\,$=$\,$0.3$. The center of $\alpha_{\rm CO}(R)$ gives 2.1\,$M_{\odot}$\,pc$^{-2}$\,(K\,km\,s$^{-1}$)$^{-1}$. The $\alpha_{\rm CO}(R)$ increases with radius, flattens out at $r=0.3\,R_{25}$, and then is fixed at 4.6\,$M_{\odot}$\,pc$^{-2}$\,(K\,km\,s$^{-1}$)$^{-1}$ (black curve in the top panel of Fig.~\ref{XCO}). Compared with the outer disk, the central $\alpha_{\rm CO}(R)$ is a factor of 2.2 (0.34\,dex) lower. The same best-fit function is used to convert the CO intensity to molecular surface density ($\Sigma_{\rm mol}$) for each galaxy, and then we calculate the molecular gas concentration ($C_{\rm mol}(\alpha_{\rm CO}[R])$) as the ratio of central molecular surface density ($\Sigma_{\rm mol}(\alpha_{\rm CO}[R]; R\leq 0.2\,R_e)$) to the disk-averaged surface density ($\Sigma_{\rm mol}(\alpha_{\rm CO}[R]; R\leq R_e)$):
\begin{equation}\label{Cmol_r}
  C_{\rm mol}(\alpha_{\rm CO}[R])=\log \frac{ \Sigma_{\rm mol}(\alpha_{\rm CO}[R]; R\leq 0.2\,R_e)  }{ \Sigma_{\rm mol}(\alpha_{\rm CO}[R]; R\leq R_e)}.
\end{equation}
\noindent
$C_{\rm mol}(\alpha_{\rm CO}[R])$ is only for understanding the effect of the same central $\alpha_{\rm CO}$ depression, not for exploring the impact of disk structure on central concentrations of molecular gas.

The degree of central depression in $\alpha_{\rm CO}$ varies from galaxy to galaxy \citep{Sandstrom2013}, imposing different level of influence on molecular gas concentrations. Considering two primary dependences of $\alpha_{\rm CO}$, \cite{Bolatto2013} suggested a prescription of $\alpha_{\rm CO}$ (their Eq.~(31)), which involves metallicity ($Z$) to take into account the CO-faint molecular gas and involves total surface density ($\Sigma_{\rm total}$) to take into account the effects of temperature and velocity dispersion. We replace the $\Sigma_{\rm total}$ with $\Sigma_{\star}$, because the stellar component is dominant in our sample. We assume that giant molecular clouds have a characteristic mean surface density of 100\,$M_{\odot}$\,pc$^{-2}$, as did \cite{Bolatto2013} to derive prediction. We therefore have
\begin{multline} \label{aco_zsig}
  \alpha_{\rm CO}(Z,\Sigma_{\star}) = 2.9 \times \exp\left(\frac{0.4}{Z/Z_{\odot}}\right)\times 
  \left( \frac{\Sigma_{\star}}{100\,M_{\odot}\,{\rm pc}^{-2}} \right)^{-\gamma} \\
  \,M_{\odot}\,{\rm pc}^{-2}\,({\rm K\,km\,s}^{-1})^{-1}, 
\end{multline}
with $\gamma$\,$=$\,$0.5$ for $\Sigma_{\star}$\,$>$\,$100\,M_{\odot}\,{\rm pc}^{-2}$ and $\gamma$\,$=$\,$0$ otherwise. $Z_{\odot}$ is the solar metallicity. Since the O$/$H measurements are not available at some locations in the galaxy, we fit a straight line to the $12+\log({\rm O}/{\rm H})$ versus $R$, and use the best-fit line to regenerate a 2D map of $12+\log({\rm O}/{\rm H})$. The solar oxygen abundance of 8.7 is adopted. The metallicity ($Z$) is given by: $\log(Z/Z_{\odot})=12+\log({\rm O}/{\rm H})-8.7$ \citep{Marino2013}. To avoid the masked region, we do azimuthal averaging of  $\Sigma_{\star}$ at each radius to obtain the radial profile of $\Sigma_{\star}$, then regenerate a 2D map of $\Sigma_{\star}$. These two maps are entered into Eq.~(\ref{aco_zsig}) to calculate $\alpha_{\rm CO}(Z,\Sigma_{\star})$. The derived $\alpha_{\rm CO}(Z,\Sigma_{\star})$ for each galaxy are marked by gray curves in Fig.~\ref{XCO}, and the average values for a given normalized radius are marked by thick black curves. The derived $\alpha_{\rm CO}(Z,\Sigma_{\star})$ are consistent with the observations in \cite{Sandstrom2013} within uncertainty. The average $\alpha_{\rm CO}(Z,\Sigma_{\star})$ yields 1.3\,$M_{\odot}$\,pc$^{-2}$\,(K\,km\,s$^{-1}$)$^{-1}$ in the center ($R$\,$=$\,0), and 5.8\,$M_{\odot}$\,pc$^{-2}$\,(K\,km\,s$^{-1}$)$^{-1}$ in the outer part (0.8\,$<$\,$R/R_{25}$\,$<$\,1). Compared with the outer disk, the central $\alpha_{\rm CO}(Z,\Sigma_{\star})$ is therefore lower by a factor of 4.5 (0.65\,dex). We convert CO intensity to molecular gas surface density using the derived $\alpha_{\rm CO}(Z,\Sigma_{\star})$ for each galaxy, and define the molecular gas concentration:
\begin{equation}\label{Cmol}
  C_{\rm mol}(\alpha_{\rm CO}[Z,\Sigma_{\star}])=\log \frac{ \Sigma_{\rm mol}(\alpha_{\rm CO}[Z,\Sigma_{\star}]; R\leq 0.2\,R_e)  }{ \Sigma_{\rm mol}(\alpha_{\rm CO}[Z,\Sigma_{\star}]; R\leq R_e)}.
\end{equation}
\noindent
The derived $C_{\rm mol}(\alpha_{\rm CO}[Z,\Sigma_{\star}])$ is used to explore the impact of disk structure on the molecular gas concentration. The calculation of molecular gas concentration (e.g., $C_{\rm mol}(\alpha_{\rm CO}[R])$ or $C_{\rm mol}(\alpha_{\rm CO}[Z,\Sigma_{\star}])$) does not depend on the normalization of $\alpha_{\rm CO}$, because any multiplicative constant term in the numerator and denominator cancels out. Instead, the shape of the $\alpha_{\rm CO}$ profile influence the molecular gas concentrations.

To estimate the uncertainty, we perform a Monte Carlo test on the $C_{\rm CO}$, $C_{\rm mol}(\alpha_{\rm CO}[R])$, and $C_{\rm mol}(\alpha_{\rm CO}[Z,\Sigma_{\star}])$, by adding random noise to CO integrated intensities, $\epsilon_{\rm disk}$, and PA$_{\rm disk}$, according to their statistical errors. We repeat the calculations with the randomly perturbed data values 1000 times and find the standard deviation of the results.

\begin{figure}
        \centering
        \includegraphics[width=0.45\textwidth]{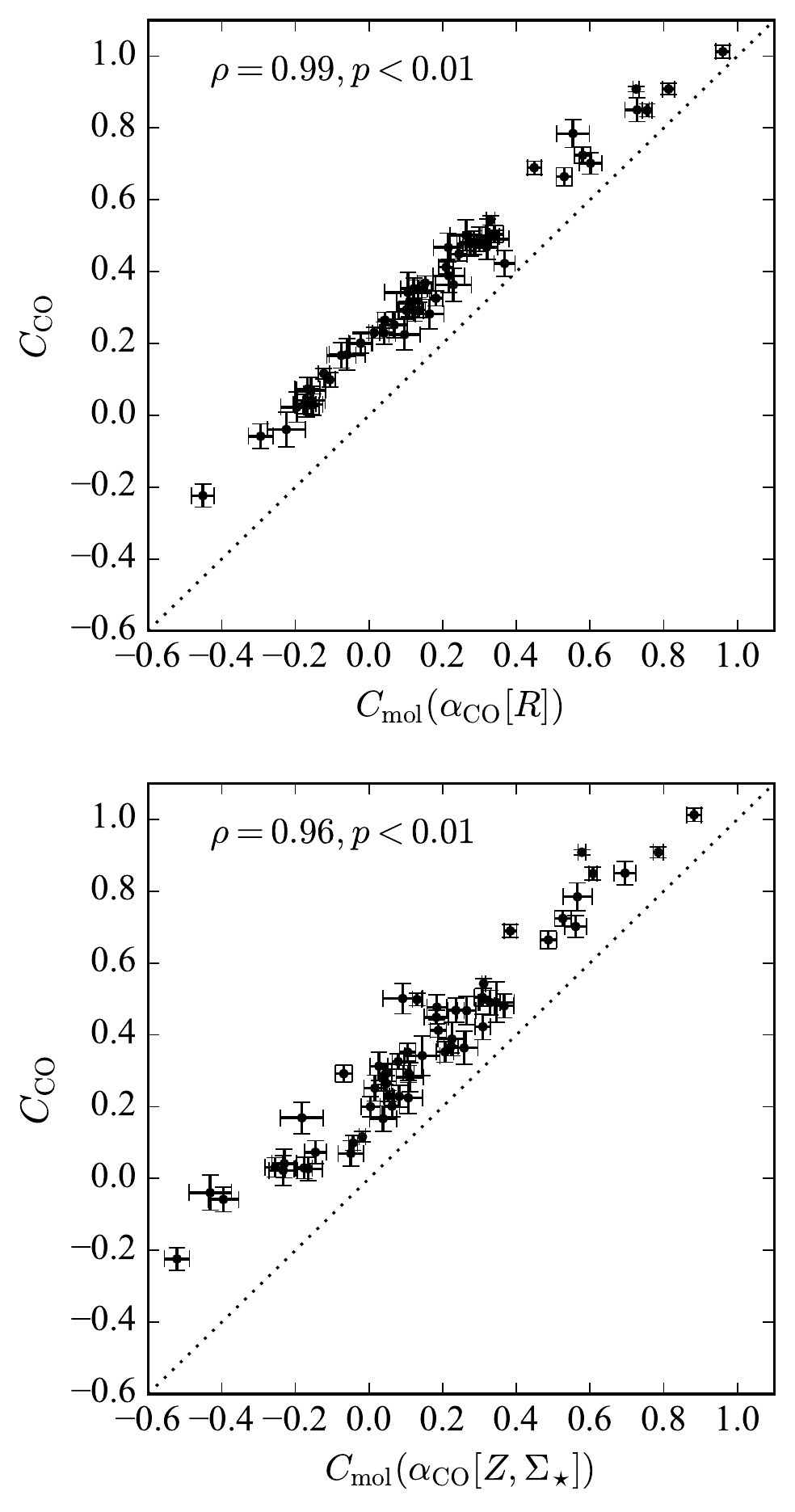}
        \caption{Correlations between CO luminosity concentration ($C_{\rm CO}$) and molecular gas mass concentration ($C_{\rm mol}$). {\it Top panel:}  $C_{\rm mol}(\alpha_{\rm CO}[R])$ derived based on $\alpha_{\rm CO}(R)$, which is a function of radius and is set the same for all the galaxies (Eq.~\ref{aco_r}). {\it Bottom panel:}  $C_{\rm mol}(\alpha_{\rm CO}[Z, \Sigma_{\star}])$  computed based on $\alpha_{\rm CO}[Z, \Sigma_{\star}]$, which is a function of $Z$ and $\Sigma_{\star}$ and varies from galaxy to galaxy (Eq.~\ref{aco_zsig}). The dashed lines mark the $1:1$ relations.
        }
        \label{Clm}
\end{figure}

 The purpose of deriving $C_{\rm mol}(\alpha_{\rm CO}[R])$ with $\alpha_{\rm CO}(R)$ fixed the same for all the galaxy is to first understand the effect of the same central $\alpha_{\rm CO}(R)$ depression, which helps reveal the effect of $\alpha_{\rm CO}(Z,\Sigma_{\star})$, which changes from galaxy to galaxy. Figure~\ref{Clm} compares $C_{\rm CO}$ with $C_{\rm mol}(\alpha_{\rm CO}[R])$ and $C_{\rm mol}(\alpha_{\rm CO}[Z,\Sigma_{\star}])$. The $C_{\rm CO}$ correlates strongly with $C_{\rm mol}(\alpha_{\rm CO}[R])$ with a Pearson correlation coefficient $\rho$\,=\,$0.99$, and $C_{\rm mol}(\alpha_{\rm CO}[R])$ are systematically 0.19\,dex lower. Similarly, in addition to being strongly correlated ($\rho$\,=\,$0.96$), $C_{\rm mol}(\alpha_{\rm CO}[Z,\Sigma_{\star}])$ are systematically 0.22\,dex lower than $C_{\rm CO}$.  The difference between $C_{\rm CO}$ and $C_{\rm mol}(\alpha_{\rm CO}[Z,\Sigma_{\star}])$ is larger than between $C_{\rm CO}$ and $C_{\rm mol}(\alpha_{\rm CO}[R])$, because the central depression is deeper in $\alpha_{\rm CO}(Z,\Sigma_{\star})$ than in $\alpha_{\rm CO}(R)$. The scatter in the $C_{\rm CO}$-$C_{\rm mol}(\alpha_{\rm CO}[R])$ relation is 0.05\,dex, which is lower than the scatter of 0.08\,dex in the $C_{\rm CO}$-$C_{\rm mol}(\alpha_{\rm CO}[Z,\Sigma_{\star}])$ relation. This is because the same $\alpha_{\rm CO}(R)$ is used, whereas $\alpha_{\rm CO}(Z,\Sigma_{\star})$ varies from galaxy to galaxy. By comparing $C_{\rm CO}$-$C_{\rm mol}(\alpha_{\rm CO}[R])$ and $C_{\rm CO}$-$C_{\rm mol}(\alpha_{\rm CO}[Z,\Sigma_{\star}])$ relation, we can be sure that the lower value in $C_{\rm mol}(\alpha_{\rm CO}[Z,\Sigma_{\star}])$ compared to $C_{\rm CO}$ is caused by the central depression in $\alpha_{\rm CO}(Z,\Sigma_{\star})$.

The caveat is that there is still a fair dispersion around the average prescription $\alpha_{\rm CO}(Z,\Sigma_{\star})$, representing the variation in local parameters such as temperature and surface density of giant molecular clouds, although $\alpha_{\rm CO}(Z,\Sigma_{\star})$ has considered the main drivers of $\alpha_{\rm CO}$ variations (see the detailed discussion in \citealt{Bolatto2013}). We use the molecular mass map derived using $\alpha_{\rm CO}(Z,\Sigma_{\star})$ and probe the connection between $C_{\rm mol}(\alpha_{\rm CO}[Z,\Sigma_{\star}])$ and disk structure. We adopt
\begin{equation}
  C_{\rm mol} \equiv C_{\rm mol}(\alpha_{\rm CO}[Z,\Sigma_{\star}])
\end{equation}
in the remainder of this paper. We have verified that the resulting $C_{\rm CO}$ and $C_{\rm mol}$ are almost identical to those computed using dilated-mask moment maps (see \citealt{Bolatto2017} for the derivation of dilated-mask moment maps), so using the dilated-mask moment maps does not affect our results. The derived $C_{\rm CO}$, $C_{\rm mol}$ and their uncertainties are listed in Table~\ref{tdata}.

%-----------------------------------------------------------------------------

\subsection{Central concentrations of star formation rate}\label{Sec_Csfr}

We use the extinction-corrected SFR derived from H$\alpha$ flux using a Salpeter initial mass function and the nebular extinction based on the Balmer decrement \citep{Bolatto2017}. SFR in pixels with EW$_{\rm H\alpha}$ less than 6 \AA\ are blanked, because they are primarily ionized by evolved stars \citep{Sanchez2014}. The derived SFR traces ongoing star formation averaged over the past $\sim$\,10\,Myr \citep{Kennicutt2012}. The SFR surface densities ($\Sigma_{\rm SFR}$) are then computed by dividing the SFR by the physical area of each pixel.

Measuring SFRs in active galaxies is challenging because tracers of young stars are to some extent contaminated with radiation from the AGN. We removed the four Seyferts and eight ambiguous galaxies, where the dominant power source may be an AGN, to avoid contamination in quantifying the SFR concentration. This leads to 45 galaxies. \cite{CidFernandes2011} showed that a value of EW$_{\rm H\alpha}$\,$\leq$\,3\,$\AA$ corresponds to retired regions in galaxies, which have ceased star formations and are now ionized by hot low-mass evolved stars. The EW$_{\rm H\alpha}$\,$\geq$\,6\,$\AA$ are primarily dominated by recent star formation, while those with 3\,$\AA$\,$<$\,${\rm EW_{H\alpha}}$\,$<$\,6\,$\AA$ are mixed \citep{Sanchez2014, Lacerda2018}. We require that the average central EW$_{\rm H\alpha}(R\leq 0.2\,R_e)$\,$\geq$\,6\,$\AA$ to ensure robust measurement of SFR. This leads to the final sample of 38 galaxies for measuring the SFR concentration.

We define the concentration of SFR ($C_{\rm SFR}$) as the ratio of the nuclear SFR surface density ($\Sigma_{\rm SFR} (R\leq 0.2\,R_e)$) to the disk-averaged SFR surface density ($\Sigma_{\rm SFR}(R\leq R_e)$):
\begin{equation}\label{dSFRavg}
  C_{\rm SFR} = \log  \frac{    \Sigma_{\rm SFR} (R\leq 0.2\,R_e) }{\Sigma_{\rm SFR}(R\leq R_e) }.
\end{equation}
\noindent
If the central SFR is enhanced with respect to global SFR, the $\Sigma_{\rm SFR} (R\leq 0.2\,R_e)$ increases for a given $\Sigma_{\rm SFR}(R\leq R_e)$, and thus $C_{\rm SFR}$ is elevated. Uncertainties of $C_{\rm SFR}$ are estimated from the errors of SFR, $\epsilon_{\rm disk}$, and PA$_{\rm disk}$ through a Monte Carlo test. The derived $C_{\rm SFR}$ and uncertainties are listed in Table~\ref{tdata}.

\section{Results}\label{results}

In this section we first demonstrate the ability of $C_{\rm CO}$ and $C_{\rm mol}$ to measure central concentrations, then explore distributions of $C_{\rm CO}$ and $C_{\rm mol}$ in barred and unbarred galaxies, and finally study the dependence of $C_{\rm CO}$ and $C_{\rm mol}$ on the strengths of disk (spiral+bar) structure. We use the Pearson correlation coefficients to analyze the relationships. The Pearson correlation coefficient measures tightness of a linear relation between two set of data. The Spearman correlation coefficient measures relation tightness, regardless of whether the relation is linear or not. Still, there is not significant difference between the two correlation coefficients of the relations presented in this paper. We use the Pearson correlation coefficient ($\rho$) in this work due to the perspicuity of linearity.

\subsection{Profiles of the molecular gas surface density}\label{SecShp}

\begin{figure}
        \centering
        \includegraphics[width=0.45\textwidth]{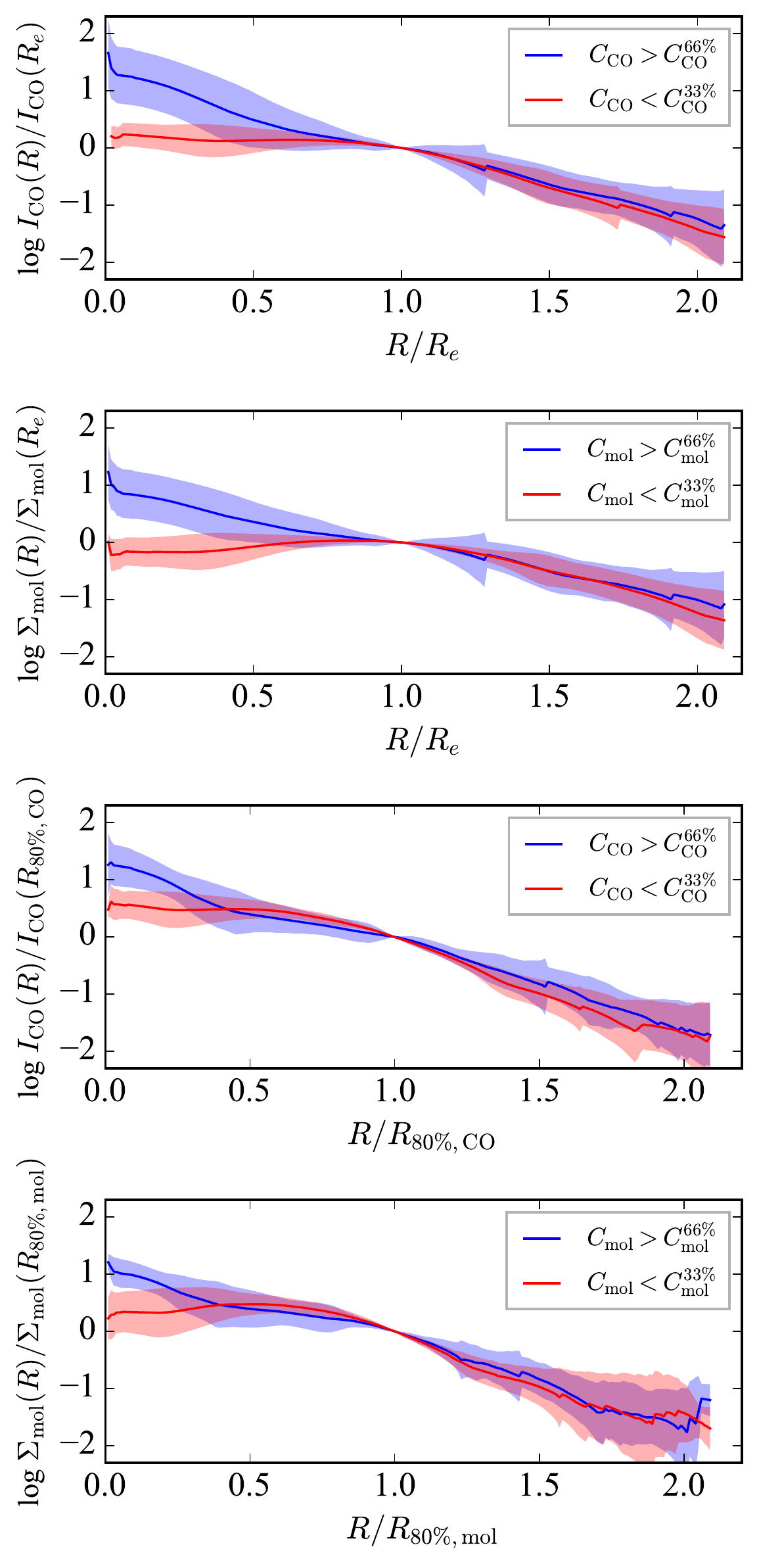}
        \caption{Average normalized profiles of CO integrated intensity ($I_{\rm CO}(R)/I_{\rm CO}(R_e)$ and $I_{\rm CO}(R)/I_{\rm CO}(R_{\rm 80\%, CO})$) and molecular mass surface density ($\Sigma_{\rm mol}(R)/\Sigma_{\rm mol}(R_e)$ and $\Sigma_{\rm mol}(R)/\Sigma_{\rm mol}(R_{\rm 80\%, mol})$). 
        Blue profiles are for high-$C_{\rm CO}$ or high-$C_{\rm mol}$ galaxies, while red profiles are for low-$C_{\rm CO}$ or low-$C_{\rm mol}$ galaxies. The shaded regions mark the standard deviation in y-axis value for a given normalized radius.
        }
        \label{shape}
\end{figure}

To demonstrate the ability of $C_{\rm CO}$ and $C_{\rm mol}$ to measure central concentrations, we extract profiles of CO intensity ($I_{\rm CO}(R)$) and molecular gas surface mass ($\Sigma_{\rm mol}(R)$; $\alpha_{\rm CO}(Z, \Sigma_{\star})$ is used) using a series of elliptical rings with measured $\epsilon_{\rm disk}$ and PA$_{\rm disk}$. $C_{\rm CO}^{33\%}=0.24$ and $C_{\rm CO}^{66\%}=0.47$ are respectively the 33th and 66th percentile of the $C_{\rm CO}$ in the sample. $C_{\rm mol}^{33\%}=0.05$ and $C_{\rm mol}^{66\%}=0.23$ are respectively the 33th and 66th percentile of the $C_{\rm mol}$ in the sample. We separate the data into four groups: high-$C_{\rm CO}$ galaxies ($C_{\rm CO}>C_{\rm CO}^{66\%}$), low-$C_{\rm CO}$ galaxies ($C_{\rm CO}<C_{\rm CO}^{33\%}$), high-$C_{\rm mol}$ galaxies ($C_{\rm mol}>C_{\rm mol}^{66\%}$), and low-$C_{\rm mol}$ galaxies ($C_{\rm mol}<C_{\rm mol}^{33\%}$).

To visualize their shape, we normalize the $I_{\rm CO}(R)$ and $\Sigma_{\rm mol}(R)$ to their values at $R$\,$=$\,$R_e$, that is, $I_{\rm CO}(R)/I_{\rm CO}(R_e)$ and $\Sigma_{\rm mol}(R)/\Sigma_{\rm mol}(R_e)$, and plot them as a function of $R/R_e$ in Fig.~\ref{shape} (top two panels). The average $I_{\rm CO}(R)/I_{\rm CO}(R_e)$ profiles of high-$C_{\rm CO}$ galaxies and average $\Sigma_{\rm mol}(R)/\Sigma_{\rm mol}(R_e)$ profiles of high-$C_{\rm mol}$ galaxies are approximately exponential, but they slightly bends upward at $R$\,$\approx$\,0.6\,$R_e$. The average profiles for low-$C_{\rm CO}$ and low-$C_{\rm mol}$ galaxies flatten below $R_e$ and become almost flat at the center. The difference between low and high concentrations is the most pronounced at the center, consistent with the definition of $C_{\rm CO}$ and $C_{\rm mol}$, which compare the value averaged within 0.2\,$R_e$ to the value averaged within $R_e$. In contrast, there is no significant difference in the profiles beyond $R_e$ between galaxies with low and high concentrations. We point out that the $C_{\rm CO}$ and $C_{\rm mol}$ reflect the global shape rather than a central precipitous rise within 0.2\,$R_e$. At the center, the $\Sigma_{\rm mol}(0)/\Sigma_{\rm mol}(R_e)$ is lower than $I_{\rm CO}(0)/I_{\rm CO}(R_e)$, because the $\alpha_{\rm CO}(Z, \Sigma_{\star})$ is on average 0.65\,dex lower in the center than in the outer disk. This is consistent with Fig.~\ref{Clm}, where the $C_{\rm mol}$ is systematically lower than $C_{\rm CO}$.

The definitions of $C_{\rm CO}$ and $C_{\rm mol}$ include a normalization term of galaxy size and therefore remove the dependence on the size. Nevertheless, we also visualize the profiles normalized by the size of the CO disk. The $R_{20\%,{\rm CO}}$ ($R_{20\%,{\rm mol}}$) and $R_{80\%,{\rm CO}}$ ($R_{80\%,{\rm mol}}$), which contains 20\% and 80\% of total CO flux (total molecular mass) are derived. We, respectively, normalize the $I_{\rm CO}(R)$ and $\Sigma_{\rm mol}(R)$ to their values at $R$\,$=$\,$R_{80\%,{\rm CO}}$ and $R$\,$=$\,$R_{80\%,{\rm mol}}$, that is, $I_{\rm CO}(R)/I_{\rm CO}(R_{80\%,{\rm CO}})$ and $\Sigma_{\rm mol}(R)/\Sigma_{\rm mol}(R_{80\%,{\rm mol}})$ (bottom two panels in Fig.~\ref{shape}).

Compared to those normalized to the values at $R$\,$=$\,$R_e$, the difference in $I_{\rm CO}(R)/I_{\rm CO}(R_{80\%,{\rm CO}})$ between high- and low-$C_{\rm CO}$ galaxies or the difference in $\Sigma_{\rm mol}(R)/\Sigma_{\rm mol}(R_{80\%,{\rm mol}})$ between high- and low-$C_{\rm mol}$ galaxies becomes smaller, especially at the center. However, other main features as in the profiles normalized using $R_e$ are still present. Inspired by the definition of galaxy light concentration \citep{Conselice2003}, we additional test two kinds of concentrations, without consideration on the stellar components, which are $\log(R_{\rm 80\%, CO}/R_{\rm 20\%, CO})$ and $\log(R_{\rm 80\%, mol}/R_{\rm 20\%, mol})$. We find that the $C_{\rm CO}$ are consistent with $\log(R_{\rm 80\%, CO}/R_{\rm 20\%, CO})$ with $\rho$\,$=$\,$0.83$ and $p$\,$<$\,$0.01$, and that $C_{\rm mol}$ are consistent with $\log(R_{\rm 80\%, mol}/R_{\rm 20\%, mol})$ with $\rho$\,$=$\,$0.73$ and $p$\,$<$\,$0.01$. We therefore confirm that the $C_{\rm CO}$ and $C_{\rm mol}$ truly reflect the central concentrations of CO intensity and molecular gas mass.

\subsection{Molecular gas concentrations in barred and unbarred galaxies}

\begin{figure}
        \centering
        \includegraphics[width=0.45\textwidth]{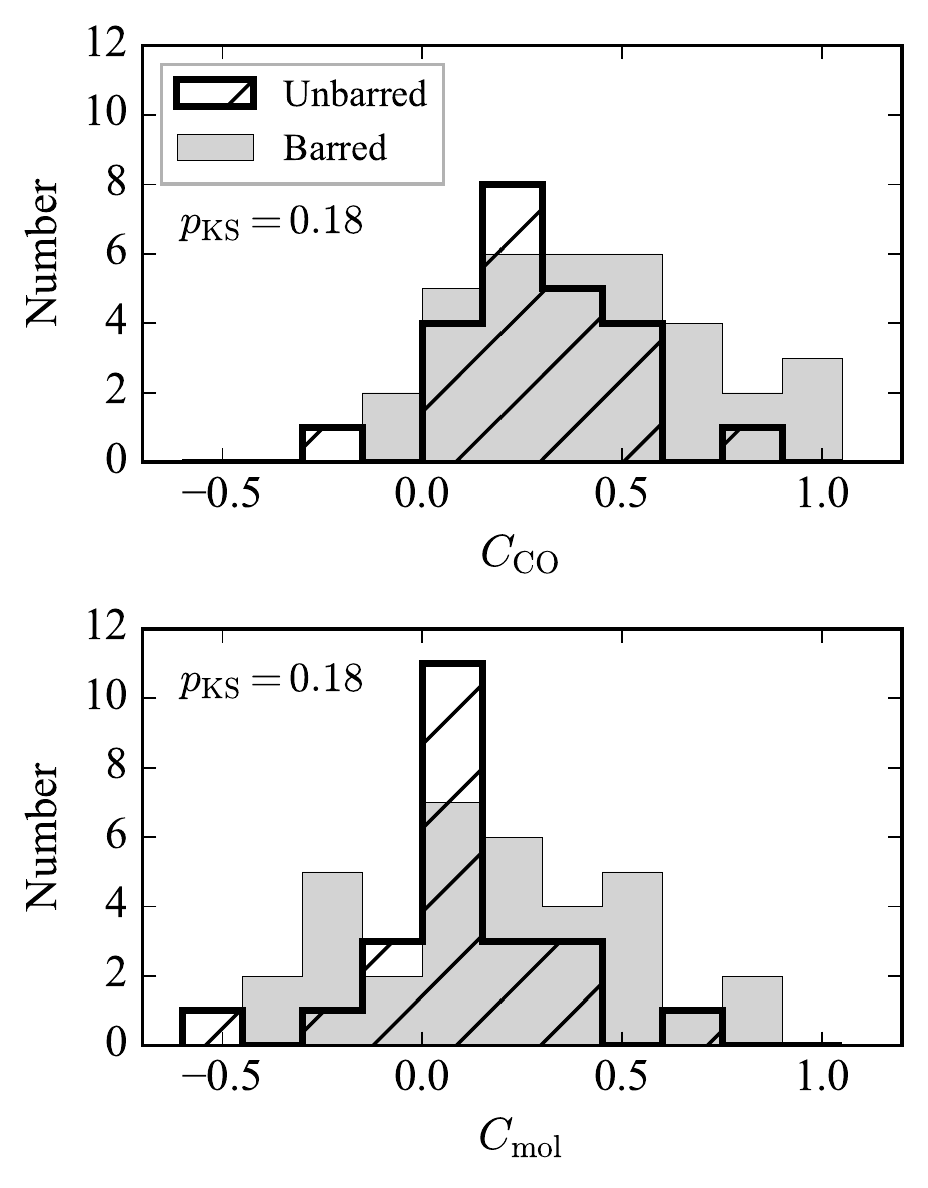}
        \caption{Histograms of the CO luminosity concentrations ($C_{\rm CO}$) and molecular gas concentrations ($C_{\rm mol}$). The results for barred and unbarred galaxies are marked in black and gray, respectively. The $p$ value of the Kolmogorov–Smirnov test ($p_{\rm KS}$) is presented. 
        }
        \label{histCC}
\end{figure}

\begin{figure}
        \centering
        \includegraphics[width=0.45\textwidth]{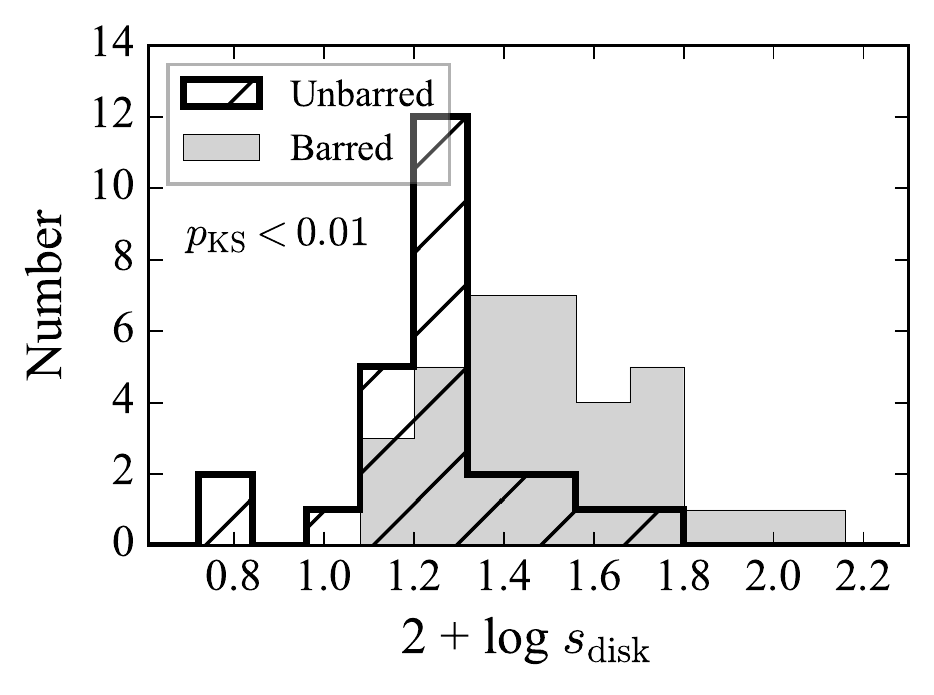}
        \caption{Histogram of the strength of non-axisymmetric disk (spiral+bar) structures ($2+\log s_{\rm disk}$).  
        The $2+\log s_{\rm disk}$ traces average spiral strength for unbarred galaxies, but traces average spiral and bar strength for barred galaxies. 
        The results for barred and unbarred galaxies are marked in black and gray, respectively. The $p$ value of the Kolmogorov–Smirnov test ($p_{\rm KS}$) is presented.
        }
        \label{histSS}
\end{figure}

Figure~\ref{histCC} presents histograms of $C_{\rm CO}$ (top) and $C_{\rm mol}$ (bottom) for unbarred and barred galaxies. Two key features are revealed. The $C_{\rm CO}$ and $C_{\rm mol}$ in unbarred galaxies are mainly moderate to low. The unbarred galaxies with the highest concentration is NGC~7819, which has $C_{\rm CO}=0.85$ and $C_{\rm mol}=0.69$, meaning that the average central CO intensity and molecular mass surface density are, respectively, a factor of 7 and 5 higher than the disk-averaged values. The concentration in NGC~7819 is comparable to the highest gas concentration of barred galaxies. In contrast, there is a wide range of $C_{\rm CO}$ and $C_{\rm mol}$ in barred galaxies. They can be as low as those in unbarred galaxies with the lowest  concentrations, or they  can reach very high values, far exceeding the concentrations in most of the unbarred galaxies. The barred galaxy with the highest concentration is NGC~447, which has $C_{\rm CO}=1.01$ and $C_{\rm mol}=0.89$, meaning that the average central CO intensity and molecular mass surface density are a factor of 10 and 8 higher than the disk-averaged values, respectively. The histogram of $C_{\rm mol}$ is similar to that of $C_{\rm CO}$, albeit shifting toward lower values.

We perform a Kolmogorov–Smirnov (KS) test for the $C_{\rm CO}$ in barred and unbarred galaxies, and a test for the $C_{\rm mol}$. We obtain the same $p_{\rm KS}$ value of 0.18 for the two tests. The $p_{\rm KS}$ is small because the barred galaxies tend to have higher $C_{\rm CO}$ or $C_{\rm mol}$, or, in other words, high $C_{\rm CO}$ or $C_{\rm mol}$ are more common in barred galaxies. But the $p_{\rm KS}$ is not small enough to reject the null hypothesis that the $C_{\rm CO}$ or $C_{\rm mol}$ in barred and unbarred galaxies are drawn from the same distribution. This is because there are a few unbarred spiral with relatively high concentrations (see also Figs.~1 and 2 in \citealt{Sheth2005}).

Assuming a constant $\alpha_{\rm CO}$, \cite{Sheth2005} showed that molecular gas concentrations in the central 1 kpc is systematically higher in barred spirals than unbarred spirals (\citealt{Sakamoto1999b}; \citealt{Kuno2007}; see also \citealt{Komugi2008}). We derived $C_{\rm mol}$ using the $\alpha_{\rm CO}(Z,\Sigma_{\star})$ that varies from galaxy to galaxy and confirmed the previous findings in the literature. The $\alpha_{\rm CO}(Z,\Sigma_{\star})$ therefore does not significantly affect the main conclusions in \cite{Sakamoto1999b}, \cite{Sheth2005}, and \cite{Kuno2007}. By comparing the molecular gas concentrations in barred and unbarred galaxies, it has been concluded that the bar is an efficient mechanism for driving gas to the galaxy center \citep{Sakamoto1999b, Sheth2005, Kuno2007}. On the other hand, this implies that the effect of spiral arms in driving gas inflow is in general moderate to weak.

Instructively, bars tend to be stronger than spiral arms by using Fourier amplitude, arm/inter-arm contrast, or gravitational torque as a measure of their strength \citep{Buta2005, Durbala2009, Bittner2017, YuHo2020}. We confirm this behavior in Fig.~\ref{histSS}, where histogram of the strengths of disk structure is shown. The disk structure in unbarred galaxies equals the spiral arms, whereas the disk structure of barred galaxy includes both spiral arms and bars. The disk structure in barred galaxies are systematically stronger than in unbarred galaxies. A KS test yields $p_{\rm KS}$\,$<$\,$0.01$,  suggesting that we can reject the null hypothesis that the two sets of $2+\log s_{\rm disk}$ are drawn from the same parent distribution. In strongly barred galaxies, the disk structure strength is driven by bars, as the strong bar is long and dominates the disk-dominated region.  In weakly barred galaxies, the disk structure strength is actually driven by spirals, as the weak bar is short \citep{Elmegreen2007} and the spirals dominate the disk-dominated region. Stronger bars tend to be associated with stronger spiral arms, either because the bars drive spirals \citep[e.g.,][]{Yuan1997, Athanassoula2009} or because the disk facilitates the formation of arms and bars to a similar extent \citep{Salo2010, Simon2019}. As in Figs.~\ref{histCC} and \ref{histSS}, barred spirals tend to have more centrally concentrated molecular gas distribution and, meanwhile, have stronger disk structures. The similarities between Figs.~\ref{histCC} and \ref{histSS} imply a connection between gas concentration and disk (spiral+bar) structure strength.

\subsection{Molecular gas concentrations and structure strength}\label{Clgtot}

\begin{figure*}
        \centering
        \includegraphics[width=0.9\textwidth]{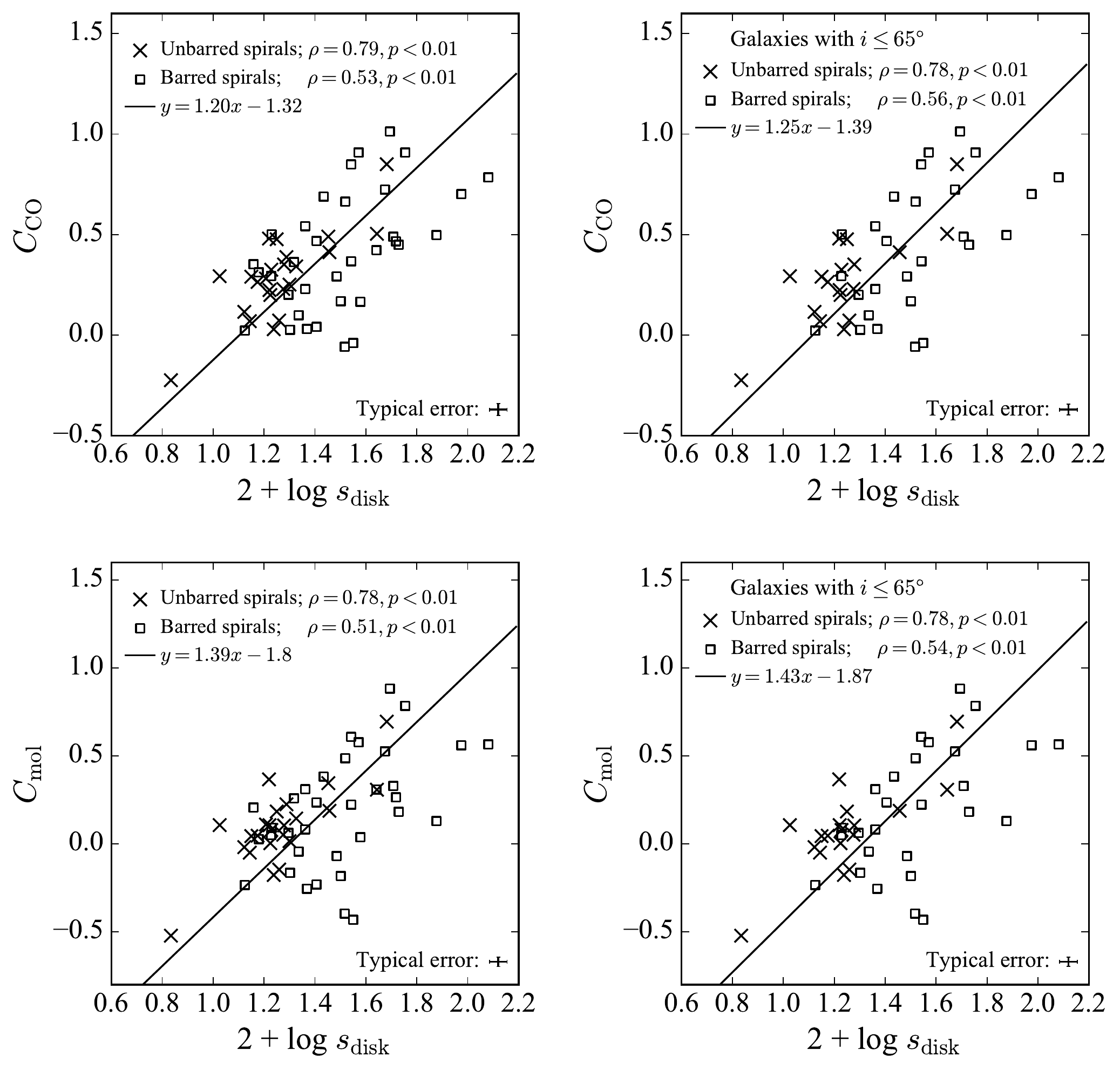}
        \caption{Dependence of CO luminosity concentrations ($C_{\rm CO}$) and molecular gas concentrations ($C_{\rm mol}$) on the strengths of disk structure ($2+\log s_{\rm disk}$). The left two panels show the results for the parent sample, whereas the right two panels show the results for galaxies with $i\leq 65\degr$. The formulas indicate linear fits considering both barred and unbarred galaxies. They are derived using PCA and marked with solid lines. The Pearson correlation coefficients, $\rho$, for unbarred and barred galaxies are given at the top of each panel. 
        }
        \label{CStr}
\end{figure*}

We study the dependence of $C_{\rm CO}$ and $C_{\rm mol}$ on the strength of non-axisymmetric disk structure using the parent sample of 57 galaxies to understand the impact of disk (spiral+bar) structure. The $C_{\rm CO}$ and $C_{\rm mol}$ increase as the disk structure strengths, regardless of whether the galaxies are barred or unbarred (left two panels in Fig.~\ref{CStr}). The Pearson correlation coefficients of $C_{\rm CO}$-$(2+\log s_{\rm disk})$ relation yields $\rho$\,$=$\,$0.79$ ($p$\,$<$\,$0.01$) and $\rho$\,$=$\,$0.53$ ($p$\,$<$\,$0.01$) for unbarred and barred galaxies, respectively. Those of $C_{\rm mol}$-$(2+\log s_{\rm disk})$ relation yields $\rho$\,$=$\,$0.78$ ($p$\,$<$\,$0.01$) and $\rho$\,$=$\,$0.51$ ($p$\,$<$\,$0.01$) for unbarred and barred galaxies, respectively. Figures~\ref{histCC} and~\ref{histSS} are projections of Fig.~\ref{CStr}. In unbarred galaxies, we reveal a novel trend that galaxies with stronger disk (spiral) structure tend to have more centrally concentrated CO or molecular gas distribution. In barred galaxies, our reported relationship is consistent with the previous study by \cite{Kuno2007}.  By assuming a constant $\alpha_{\rm CO}$ and using gravitational torque as a measure of bar strength calculated from \cite{Laurikainen2002}, \cite{Kuno2007} presented a fairly tight positive correlation between molecular gas concentration in the central $\sim$2\,kpc and bar strength.

The relations for unbarred and barred galaxies are similar (Fig.~\ref{CStr}), although unbarred galaxies tend to have lower gas concentration and weaker structure compared to barred galaxies (Fig.~\ref{histCC} and \ref{histSS}).  It may imply that the relations of gas concentration against structure strength for unbarred and barred galaxies are drawn from a unified correlation. We derive best-fit straight lines for the correlation considering both unbarred and barred spirals using principal component analysis \citep[PCA;][]{Pearson1901}. Unlike the least squares fit, which only considers the deviation of $y$-axis values from a model, the PCA fit simultaneously takes into account the $x$- and $y$-axis values. The PCA fit is preferred in this work because there should be some degree of intrinsic dispersion in the $x$-axis values of the $C_{\rm mol}$- or $C_{\rm CO}$-$(2+\log s_{\rm disk})$ relations. PCA transforms the data set, for example, $C_{\rm CO}$ versus $2+\log s_{\rm disk}$, into a set of coefficients for two new orthonormal bases (which replace the old bases: unit $C_{\rm CO}$ and $2+\log s_{\rm disk}$ along the $\vec{x}$ and $\vec{y}$ direction, respectively.). The two new orthonormal bases are the eigenvectors of the data covariant matrix and are ordered according to their eigenvalues to successively maximize variance of the data. The slope of the straight line is determined by the direction of the eigenvector with the largest eigenvalue, along which the data show the greatest variance. The straight line is the best-fit linear relation between the two parameters. The best-fit line for the $C_{\rm CO}$-$(2+\log s_{\rm disk})$ relation yields
\begin{equation}\label{cos}
  C_{\rm CO} = (1.20\pm0.24) \times (2+\log s_{\rm disk}) - (1.32\pm0.34).
\end{equation}
The uncertainties are obtained by repeating 1000 times PCA using samples of the same size, regenerated using bootstrapping with replacement, and computing standard deviations of the results. The scatter in $C_{\rm CO}$ for a given $2+\log s_{\rm disk}$ is 0.24\,dex. The best-fit line for the $C_{\rm mol}$-$(2+\log s_{\rm disk})$ relation yields
\begin{equation}\label{mols}
  C_{\rm mol} = (1.39\pm0.33) \times (2+\log s_{\rm disk}) - (1.80\pm0.47).
\end{equation}
The scatter in $C_{\rm mol}$ for a given $2+\log s_{\rm disk}$ is 0.29\,dex. The slopes of the $C_{\rm CO}$-$(2+\log s_{\rm disk})$ and $C_{\rm mol}$-$(2+\log s_{\rm disk})$ relations are consistent with each other within uncertainty. The intercept of the $C_{\rm mol}$-$(2+\log s_{\rm disk})$ relation is lower then the intercept of the $C_{\rm CO}$-$(2+\log s_{\rm disk})$ relation with statistical significance, because $C_{\rm mol}$ is systematically lower than $C_{\rm CO}$ (Fig.~\ref{Clm}).

Although we have used an elliptical aperture, with measured $\epsilon_{\rm disk}$ and PA$_{\rm disk}$, to account for the inclination effect, we are not able to completely eliminate this effect if the inclination angle ($i$) is too high. Pixels along the minor axis of the galaxy observed CO emission from a larger range of radius than those along the major axis. This effect becomes more significant as $i$ increases, making $C_{\rm CO}$ or $C_{\rm mol}$ in galaxies with high $i$ more uncertain. The inclination effect has been mitigated because edge-on galaxies have been removed during the sample selection (Sect.~\ref{data}). The measurement of $2+\log s_{\rm disk}$ is less affected as the resolution in SDSS image is better. 12 galaxies in our sample are highly inclined ($65\degr$\,$<$\,$i$\,$<$\,$75\degr$). In Fig.~\ref{CStr} (right two panels), we plot the relationships for a subsample of 45 galaxies with $i$\,$\leq$\,$65\degr$. The new $C_{\rm CO}$- and $C_{\rm mol}$-$(2+\log s_{\rm disk})$ relations for barred and unbarred galaxies remains almost unchanged.

The best-fit line derived for the $C_{\rm CO}$-$(2+\log s_{\rm disk})$ relation considering both the unbarred and barred galaxies with $i \leq 65\degr$ yields
\begin{equation}\label{cosi}
  C_{\rm CO} = (1.25\pm0.26) \times (2+\log s_{\rm disk}) - (1.39\pm0.36) \text{~~~for $i\leq 65\degr$}.
\end{equation}
The scatter in $C_{\rm CO}$ for a given structure strength is 0.26\,dex.
The same for the $C_{\rm mol}$-$(2+\log s_{\rm disk})$ relation yields
\begin{equation}\label{molsi}
  C_{\rm mol} = (1.43\pm0.37) \times (2+\log s_{\rm disk}) - (1.87\pm0.52) \text{~~~for $i\leq 65\degr$}.
\end{equation}
The scatter in $C_{\rm mol}$ for a given structure strength is 0.31\,dex. The best-fit lines (Eqs. \ref{cosi} and \ref{molsi}) are consistent with the results considering all galaxies (Eqs. \ref{cos} and \ref{mols}) within uncertainty. These results show that although the inclination affects the calculated gas concentrations in fact, this effect is not significant in our sample.

\subsection{Uncertainty from the bar identification}\label{barid}
%%%%%%%%%%%%%%%%%%%%%%%%%%%%%%%%%%%%%%%%%%%%%%%%%%%%%%%%%%%%%
\begin{table*}
\caption{\label{tid}Ordinary correlation coefficients.
% Column (1): the strength of non-axisymmetric disk structure ($2+\log s_{\rm disk}$); column (2): the central concentration of CO luminosity ($C_{\rm CO}$) or molecular gas ($C_{\rm mol}$); column (3): subsample of unbarred or barred galaxies; column (4): correlation coefficients calculated using bar classification from \cite{Walcher2014}; column (5): correlation coefficients calculated using bar classification based on isophotes (Section~\ref{Methods}); column (6): correlation coefficients calculated using bar classification from \cite{JMA2017}. 
}
\renewcommand\arraystretch{1.65}
\centering
\begin{tabular}{c|c|c|c|c|c|c}
\hline
Row & Par. 1 & Par. 2 & Barred or & Corr. coeff. with bar classif. & Corr. coeff. with bar classif. & Corr. coeff. with bar classif.     \\
  &  &        &  Unbarred&    from Walcher+2014    &  from isophotal analysis & from Méndez-Abreu+2017    \\
%\hline
 (1)   & (2) & (3)         & (4)      & (5) & (6)  & (7) \\
\hline
\multirow{2}*{[1]} & \multirow{2}*{$2+\log\,s_{\rm disk}$} &  \multirow{2}*{$C_{\rm CO}$} &  Unbarred & $\rho=0.79$, $p<0.01$   & $\rho=0.64$, $p<0.01$ & $\rho=0.66$, $p<0.01$  \\
\cline{4-7}
 &   &   & Barred & $\rho=0.53$, $p<0.01$   & $\rho=0.45$, $p=0.02$ & $\rho=0.55$, $p<0.01$ \\
\hline
\multirow{2}*{[2]} & \multirow{2}*{$2+\log\,s_{\rm disk}$} &  \multirow{2}*{$C_{\rm mol}$} &  Unbarred & $\rho=0.78$, $p<0.01$   & $\rho=0.61$, $p<0.01$ & $\rho=0.62$, $p<0.01$  \\
\cline{4-7}
 & &   & Barred & $\rho=0.51$, $p<0.01$   & $\rho=0.49$, $p<0.01$ & $\rho=0.56$, $p<0.01$  \\
\hline
\multirow{2}*{[3]} &\multirow{2}*{$2+\log\,s_{\rm disk}$} &  \multirow{2}*{$C_{\rm SFR}$} &  Unbarred & $\rho=0.71$, $p<0.01$   & $\rho=0.61$, $p<0.01$ & $\rho=0.59$, $p<0.01$  \\
\cline{4-7}
 && &  Barred &$\rho=0.70$, $p<0.01$   & $\rho=0.73$, $p<0.01$ & $\rho=0.75$, $p<0.01$  \\
\hline
\end{tabular}
%\tablefoot{
%}
\end{table*}
%%%%%%%%%%%%%%%%%%%%%%%%%%%%%%%%%%%%%%%%%%%%%%%%%%%%%%%%%%%%%

Our results for barred and unbarred galaxies are obtained using bar identification based on visual inspection from \cite{Walcher2014}. Accordingly, 34 (60\%) galaxies are barred galaxies and 23 (40\%) galaxies are unbarred galaxies. The bar fraction is consistent with previous studies that nearly 60\% of disk galaxies in the local universe have a bar \cite[e.g.,][]{Sheth2008, Aguerri2009, Simon2016, Erwin2018}. Still, there is uncertainty in identifying bars and it may influence our results. To understand the uncertainty and make our results more robust, we use two more independent bar identifications.  The first independent bar identification is based on isophotal analysis (Sect.~\ref{ISO}). Isophotes are widely adopted to identify and quantify bars \citep[e.g.,][]{Athanassoula2002, Laine2002, Erwin2003, Menendez-Delmestre2007, Aguerri2009, Li2011}. As a result, there are 25 (44\%) barred galaxies and 32 (56\%) unbarred galaxies. This derived bar fraction is underestimated, likely because the isophote-based approach is less sensitive to weak bars. The short and weak bars observed in SDSS images may not have an apparent effect on the isophotes due to limited image resolution \citep{Erwin2018}. Another independent result is from \cite{JMA2017}, who performed 2D multicomponent decomposition for the CALIFA galaxies. A galaxy is considered to be a barred galaxy if a bar is included in the decomposition. The inclusion of a bar model was determined by iteratively fitting different composite models and finding the one that best matches the observed structural properties \citep{JMA2017}. There are 32 (56\%) barred galaxies and 25 (44\%) unbarred galaxies.

We calculate $\rho$ for the $C_{\rm CO}$- and $C_{\rm mol}$-$(2+\log s_{\rm disk})$ correlations in barred and unbarred galaxies using bar identification based on isophotal analysis and the 2D decomposition (rows [1] and [2], respectively, in Table~\ref{tid}). If the two new bar identifications are used, the $\rho$ for $C_{\rm CO}$- and $C_{\rm mol}$-$(2+\log s_{\rm disk})$ relations in unbarred galaxies ranges from 0.61 to 0.66 ($p$\,$<$\,$0.01$). Likewise, the $\rho$ for the relations in barred galaxies ranges from 0.45 to 0.56 ($p$\,$<$\,$0.01$). Overall, the change in $\rho$ is small and each correlation remains statistical significant ($p$\,$<$\,$0.05$). The three independent bar identifications yield consistent results. The uncertainty from the bar identification therefore does not adversely affect our results.

\subsection{Partial correlation coefficients}
%%%%%%%%%%%%%%%%%%%%%%%%%%%%%%%%%%%%%%%%%%%%%%%%%%%
\begin{table*}
\caption{\label{trho}Partial correlation coefficients. 
%Column (1): the strength of non-axisymmetric disk structure ($2+\log\,s_{\rm disk}$); Column (2):  central concentrations of CO luminosity ($C_{\rm CO}$), molecular gas ($C_{\rm mol}$), and star formation ($C_{\rm SFR}$); Column (3): bar classification from \cite{Walcher2014}; Column (4): ordinary correlation coefficients; Column (5): mutual dependence to be remove; Column (6): partial correlation coefficients; Column (7): the $p$ value for testing non-correlation.
}
\renewcommand\arraystretch{1.65}
\centering
\begin{tabular}{c|c|c|c|c|c|c}
\hline
Row & Par. 1 & Par. 2 & Barred or & Ordinary & Dependence & Partial  \\ %& $p$   \\
 &   &        &  unbarred&  Corr. Coeff.      & removed &  Corr. Coeff.  \\ % &  \\
%\hline
 (1)   & (2) & (3)         & (4)    & (5)  & (6) & (7) \\ %& (8) \\
\hline
\multirow{2}*{[1]} & \multirow{2}*{$2+\log\,s_{\rm disk}$} &  \multirow{2}*{$C_{\rm CO}$} &  Unbarred & $\rho=0.79,p<0.01$ & \multirow{2}*{$B/T$}  & $\rho^\prime=0.79,p<0.01$ \\% & $<0.01$ \\
\cline{4-5} \cline{7-7}
  & &   & Barred & $\rho=0.53, p<0.01$ & & $\rho^\prime=0.52,p=0.01$ \\% & 0.02 \\
\hline
\multirow{2}*{[2]} & \multirow{2}*{$2+\log\,s_{\rm disk}$} &  \multirow{2}*{$C_{\rm mol}$} &  Unbarred & $\rho=0.78,p<0.01$ & \multirow{2}*{$B/T$} & $\rho^\prime=0.78,p<0.01$ \\% & $<0.01$   \\
\cline{4-5} \cline{7-7}
 & &  & Barred & $\rho=0.51,p<0.01$ & & $\rho^\prime=0.50,p=0.01$ \\% & 0.02  \\
\hline
\multirow{2}*{[3]} & \multirow{2}*{$2+\log\,s_{\rm disk}$} &  \multirow{2}*{$C_{\rm SFR}$} &  Unbarred & $\rho=0.71,p<0.01$ & \multirow{2}*{$B/T$} & $\rho^\prime=0.62, p<0.01$ \\%  & $<0.01$   \\
\cline{4-5} \cline{7-7}
& & & Barred & $\rho=0.70,p<0.01$ &  & $\rho^\prime=0.70,p<0.01$ \\%  & $<0.01$   \\
\hline
%[6] & $2+\log\,s_{\rm disk}$ & $C_{\rm SFR}$ & Both & &\\
\multirow{2}*{[4]} & \multirow{2}*{$2+\log\,s_{\rm disk}$} &  \multirow{2}*{$C_{\rm SFR}$} &  Unbarred & $\rho=0.71,p<0.01$ & \multirow{2}*{$C_{\rm CO}$} & $\rho^\prime=-0.16, p=0.48$ \\% & $0.61$   \\
\cline{4-5} \cline{7-7}
& & & Barred & $\rho=0.70,p<0.01$ &  & $\rho^\prime=0.14,p=0.45$ \\%  & $0.50$   \\
\hline
\multirow{2}*{[5]} & \multirow{2}*{$2+\log\,s_{\rm disk}$} &  \multirow{2}*{$C_{\rm SFR}$} &  Unbarred & $\rho=0.71,p<0.01$ & \multirow{2}*{$C_{\rm mol}$} & $\rho^\prime=-0.18,p=0.45$ \\%  & $0.51$   \\
\cline{4-5} \cline{7-7}
& & & Barred & $\rho=0.70,p<0.01$ &  & $\rho^\prime =0.09,p=0.63$ \\%  & $0.70$  \\
\hline
\end{tabular}
%\tablefoot{Column (1): strength of non-axisymmetric disk structure. It is average spiral strength for unbarred galaxies but average strength of spiral and bar for barred galaxies. Column (2): molecular gas concentrations or central enhancement of central star formation. Column (3): ordinary Pearson correlation coefficient between param. 1 and 2. Column (4): galaxy light concentration index measured at {\it r} band. It is the mutual dependence to be removed. Column (5): partial correlation coefficient between param. 1 and 2 with the mutual dependence on $C_{r\text{-band}}$ removed. Column (6): the $p$ value for testing non-correlation. Column (7): sample for the calculation.
%}
\end{table*}

%%%%%%%%%%%%%%%%%%%%%%%%%%%%%%%%%%%%%%%%%%%%%%%%%%

\begin{figure}
        \centering
        \includegraphics[width=0.48\textwidth]{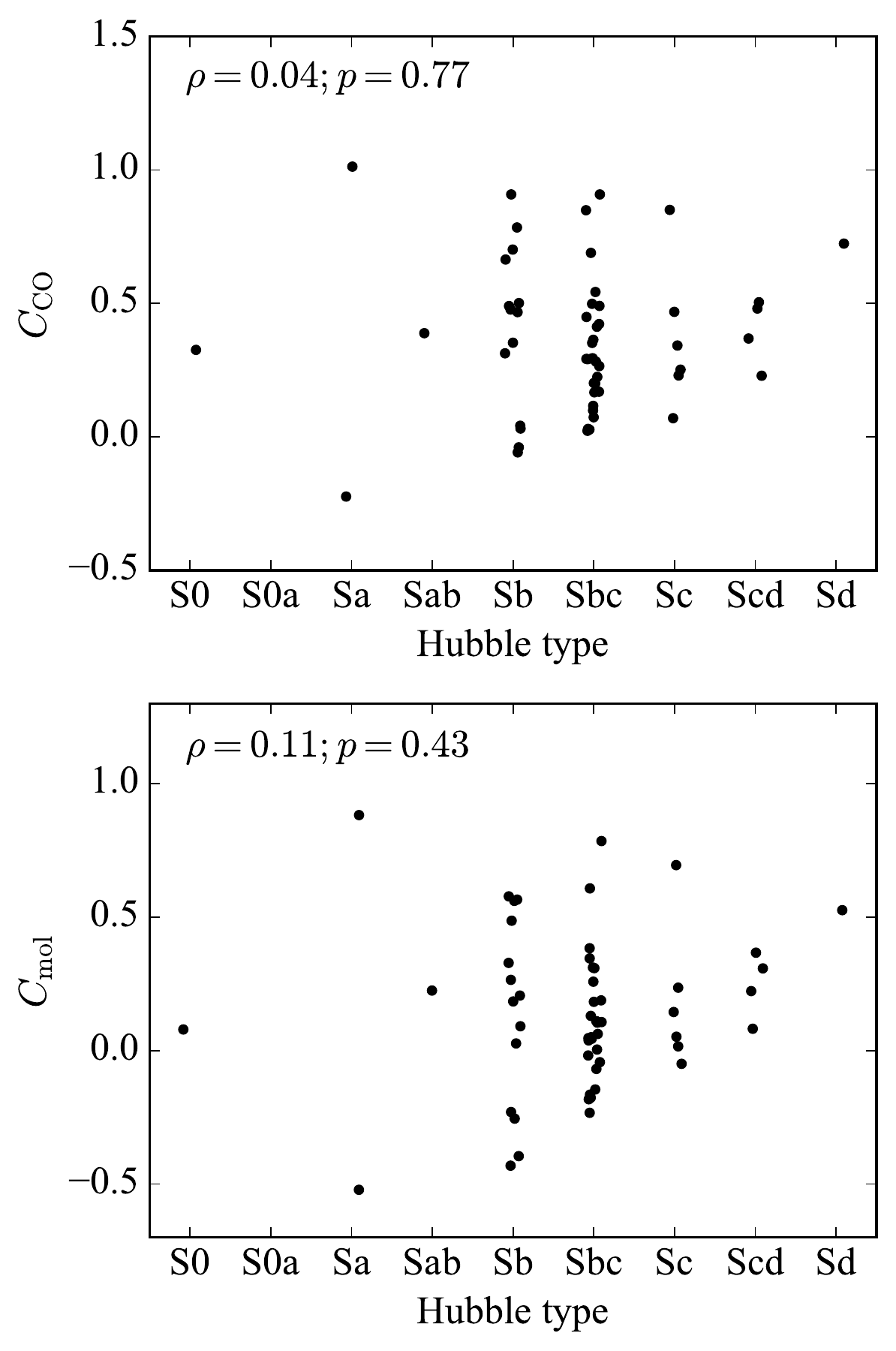}
        \caption{Comparison of $C_{\rm CO}$ and $C_{\rm mol}$ with Hubble type. Black points show the data.       The Hubble types are from \cite{Walcher2014}, and the corresponding numerical Hubble stages are used to calculate the Pearson correlation coefficients between the two parameters. 
        }
        \label{Hubble_C}
\end{figure}

In Fig.~\ref{CStr} we show that galaxies with stronger disk (spiral+bar) structure tend to have more centrally concentrated distribution of CO intensity and molecular gas mass. However, there may be a network of interdependences between gas concentrations, disk structure, and bulge properties. \cite{Komugi2008} found that early-type galaxies show higher CO luminosity concentration, defined as a ratio of average CO intensity within central 1\,kpc and 3\,kpc, and these authors attributed the trend to the larger bulges seen in earlier Hubble types. However, their definition of concentration used a fixed aperture and had a dependence on galaxy size, as early-type galaxies are larger \citep[e.g.][]{MunozMateos2015}. By using galaxy size as aperture, \cite{Kuno2007} found no link between gas concentrations and Hubble types. Consistent with \cite{Kuno2007}, we find that the correlation between $C_{\rm CO}$ and Hubble types ($\rho$\,$=$\,$0.04$ and $p$\,$=$\,$0.77$; see Fig.~\ref{Hubble_C}) and the correlation between $C_{\rm mol}$ and Hubble types ($\rho$\,$=$\,$0.11$ and $p$\,$=$\,$0.43$) are not apparent.

We note that the Hubble classification system is not only based on the visual inspection of relative bulge size, but also the character and apparent resolution of spiral arms \citep{Hubble1926, Sandage1975}. In addition to larger bulges, early type galaxies tend to have a stronger bar \citep{Gadotti2011}, and more tightly wound \citep{Yu2019} and weaker \citep{YuHo2020} spiral arms. The Hubble type may not reveal any unique galaxy properties. We therefore do not apply the analysis of partial correlation coefficients described below to the Hubble types.

One possible concern is that the central bulge may adversely affect the reported relationships, or drive them in some way. Both spiral arms and bars may be associated with the bulges. Bars would lose angular momentum to the disk, bulge, and halo during the secular evolution \citep{Tremaine1984, Sellwood1980, Debattista1998, Athanassoula2003}, and thus slow down, which makes the star obits in the bar more elongated. The bar become longer and, meanwhile, stronger as the stars are more concentrated in the bar \citep{Elmegreen2007}. A more prominent bulge indicates a more evolved system and thus a more evolved bar, resulting in an increase in bar strengths \citep{Laurikainen2007, Zhou2015, Gadotti2011}.  A massive classical bulge weakens spiral arms \citep{Bittner2017, YuHo2020}, as massive bulges would stabilize the disk and therefore reduce the dynamically active disk mass that responds to the spiral perturbation \citep{Bertin1989}. In addition to the connection to classical bulges, bars and spirals contribute to the growth of central disky pseudo bulges \citep{Kormendy2004}. We use the bulge-to-total light ratio ($B/T$) from \cite{JMA2017} as a proxy to gauge the degree of bulge prominence. To understand the possible influence from the bulge, we compute partial Pearson correlation coefficients ($\rho^{\prime}$) by removing common dependence on $B/T$ using the python package {\tt pingouin} \citep{Vallat2018} (rows [1] and [2] in Table~\ref{trho}). In both barred and unbarred galaxies, the $\rho^{\prime}$ of $C_{\rm CO}$- and $C_{\rm mol}$-$(2+\log s_{\rm disk})$ relations remain almost unchanged compared to the original ordinary Pearson correlation coefficients $\rho$. It suggests that our results are not induced by the central bulges.

%------------------------------------------------------------------------------
%------------------------------------------------------------------------------

\section{Discussion}\label{discussion}

\begin{figure*}
        \centering
        \includegraphics[width=0.95\textwidth]{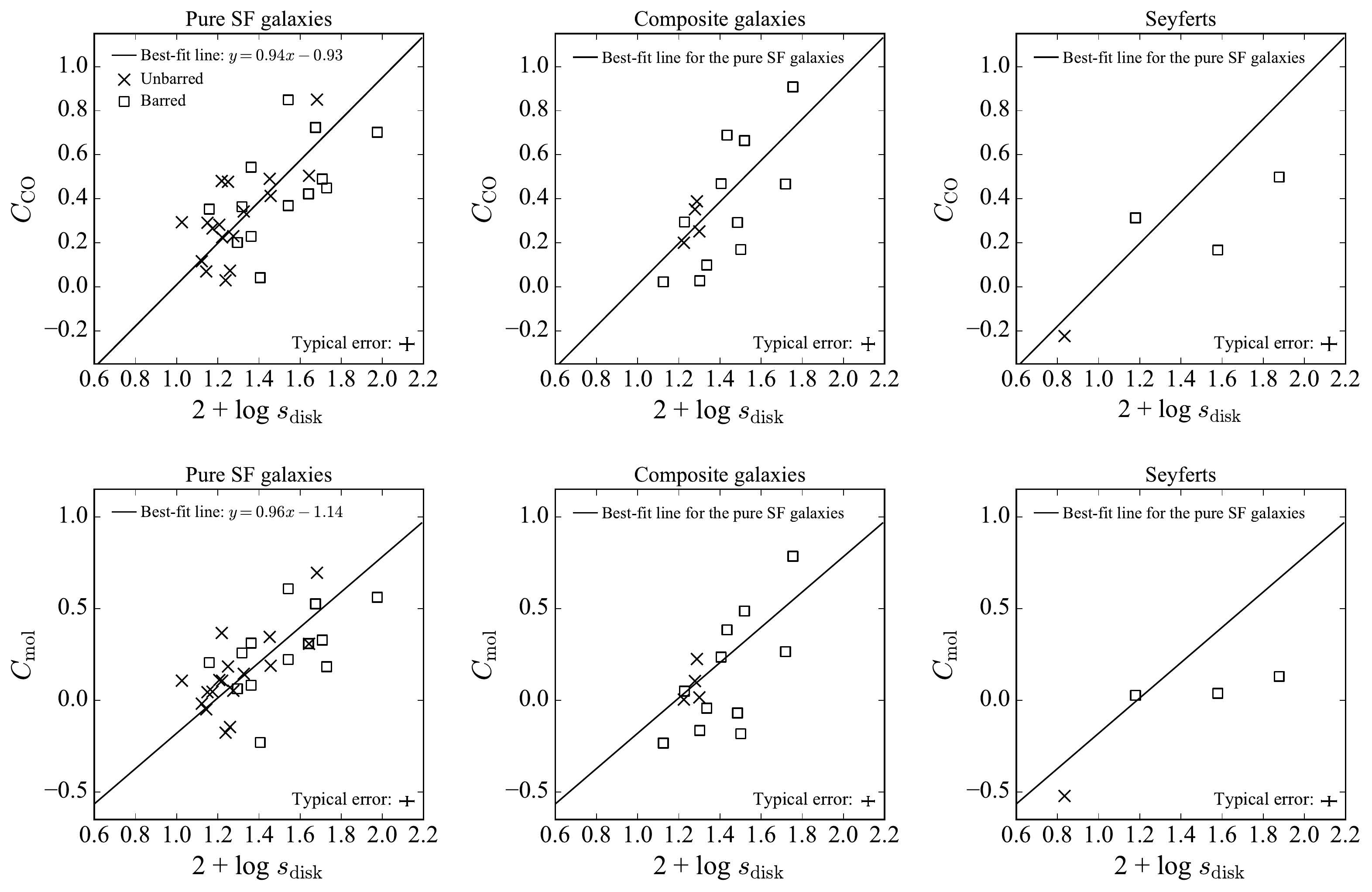}
        \caption{Dependence of $C_{\rm CO}$ and $C_{\rm mol}$ on $2+\log s_{\rm disk}$ for pure SF, composite, and Seyfert galaxies. The solid line marks the best-fit linear function of the relation constructed using the sample of pure SF galaxies.
        }
        \label{SFSey}
\end{figure*}

\begin{figure}
        \centering
        \includegraphics[width=0.45\textwidth]{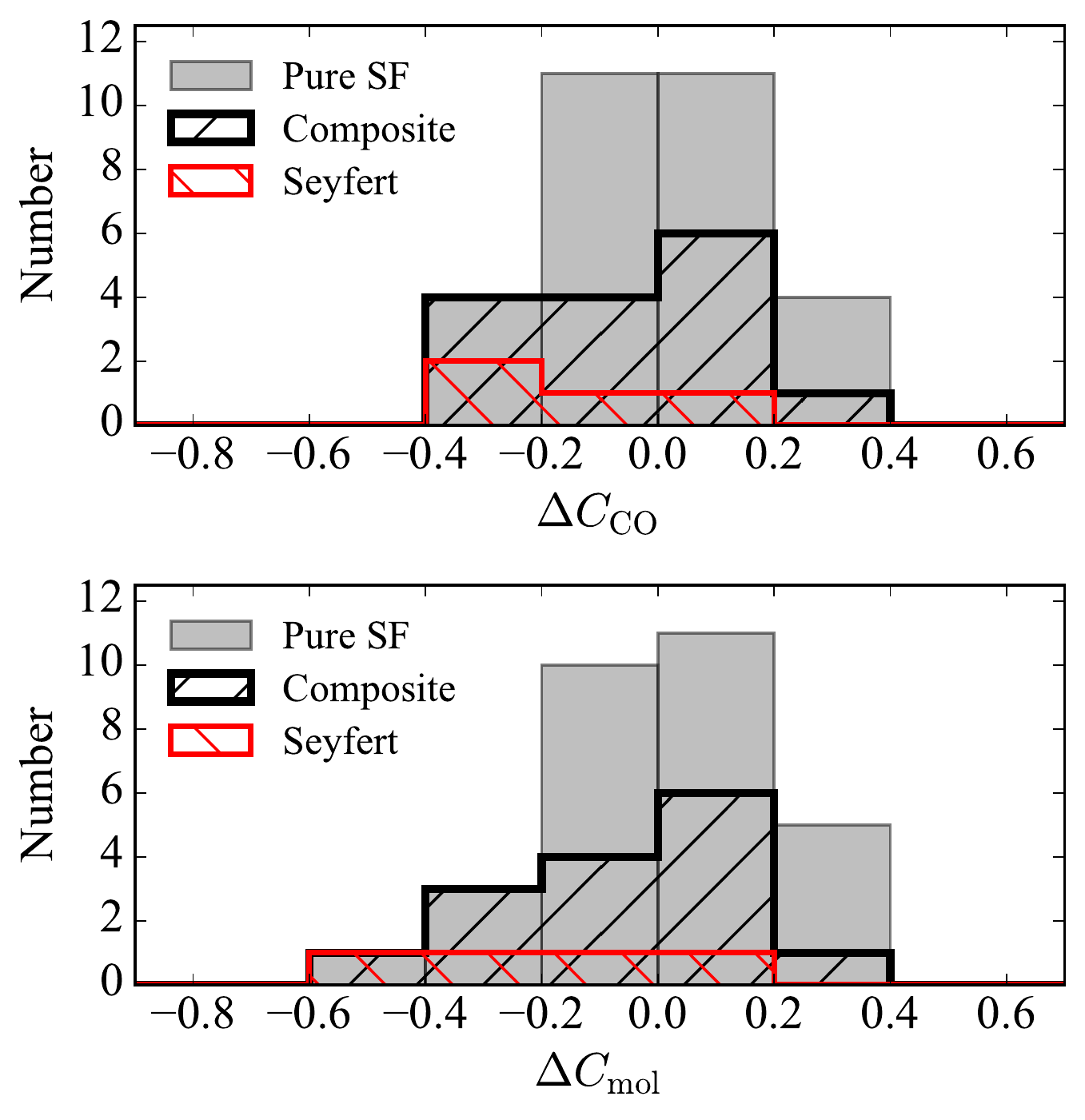}
        \caption{Histogram of $\Delta C_{\rm CO}$ and $\Delta C_{\rm mol}$ for pure SF, composite, and Seyfert galaxies.    $\Delta C_{\rm CO}$ ($\Delta C_{\rm mol}$) is the difference between the measured $C_{\rm CO}$ ($C_{\rm mol}$) and, at the measured strength, the value of the best-fit $C_{\rm CO}$-$(2+\log s_{\rm disk})$ ($C_{\rm mol}$-$(2+\log s_{\rm disk})$) relation constructed using pure SF galaxies as shown in Fig.~\ref{SFSey}.
        }
        \label{del_C}
\end{figure}

%--------------------------------------------------------------------
\begin{figure}
        \centering
        \includegraphics[width=0.4\textwidth]{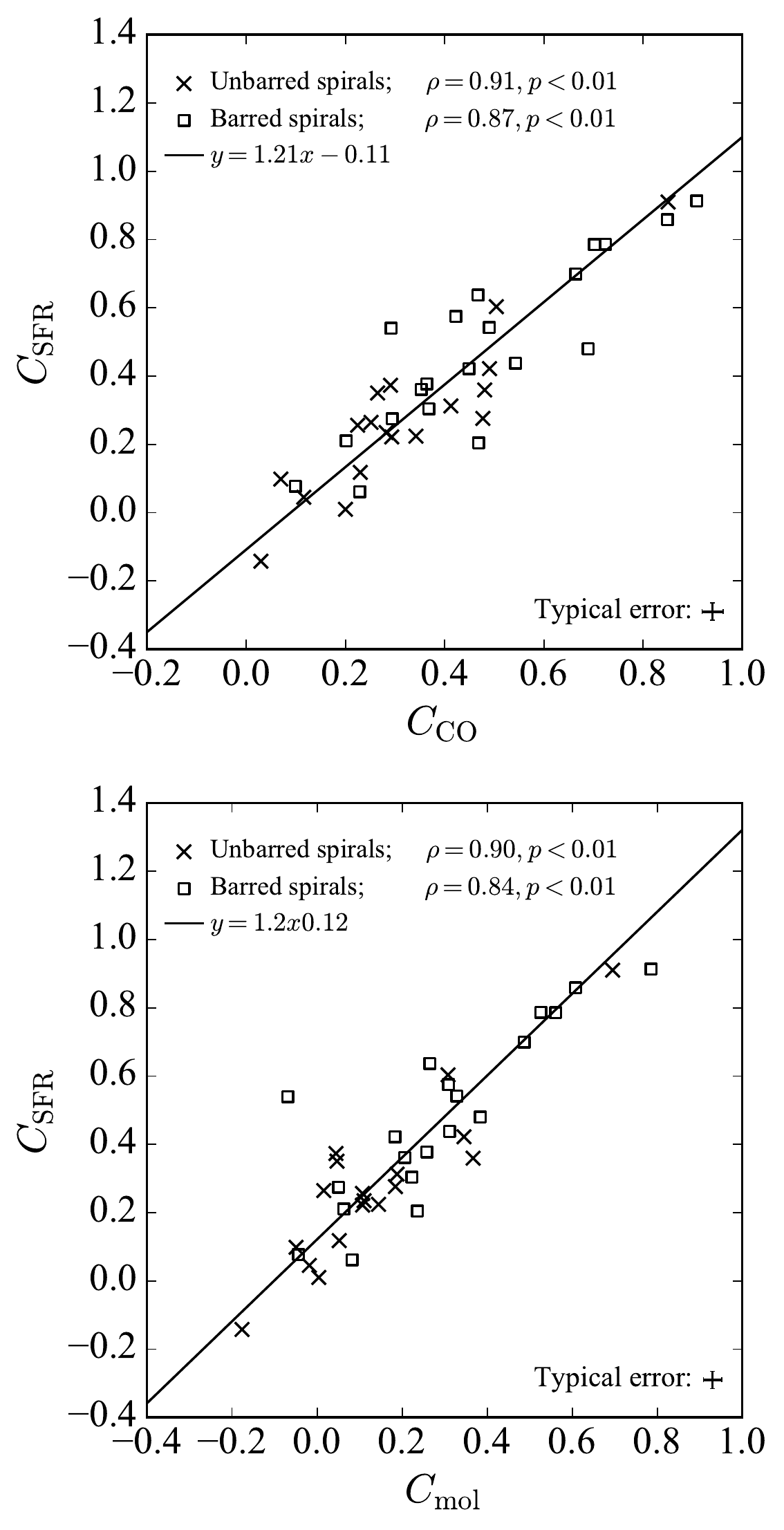}
        \caption{Dependence of $C_{\rm SFR}$ on $C_{\rm CO}$ and $C_{\rm mol}$. AGNs, LINERs and galaxies with a quenched center have been excluded. The Pearson correlation coefficients, $\rho$, for unbarred, barred, and all galaxies are given at the top of each panel. The formulas indicate the linear fits, derived using PCA and marked with solid lines.  
        }
        \label{dsfr_cog}
\end{figure}

\begin{figure}
        \centering
        \includegraphics[width=0.4\textwidth]{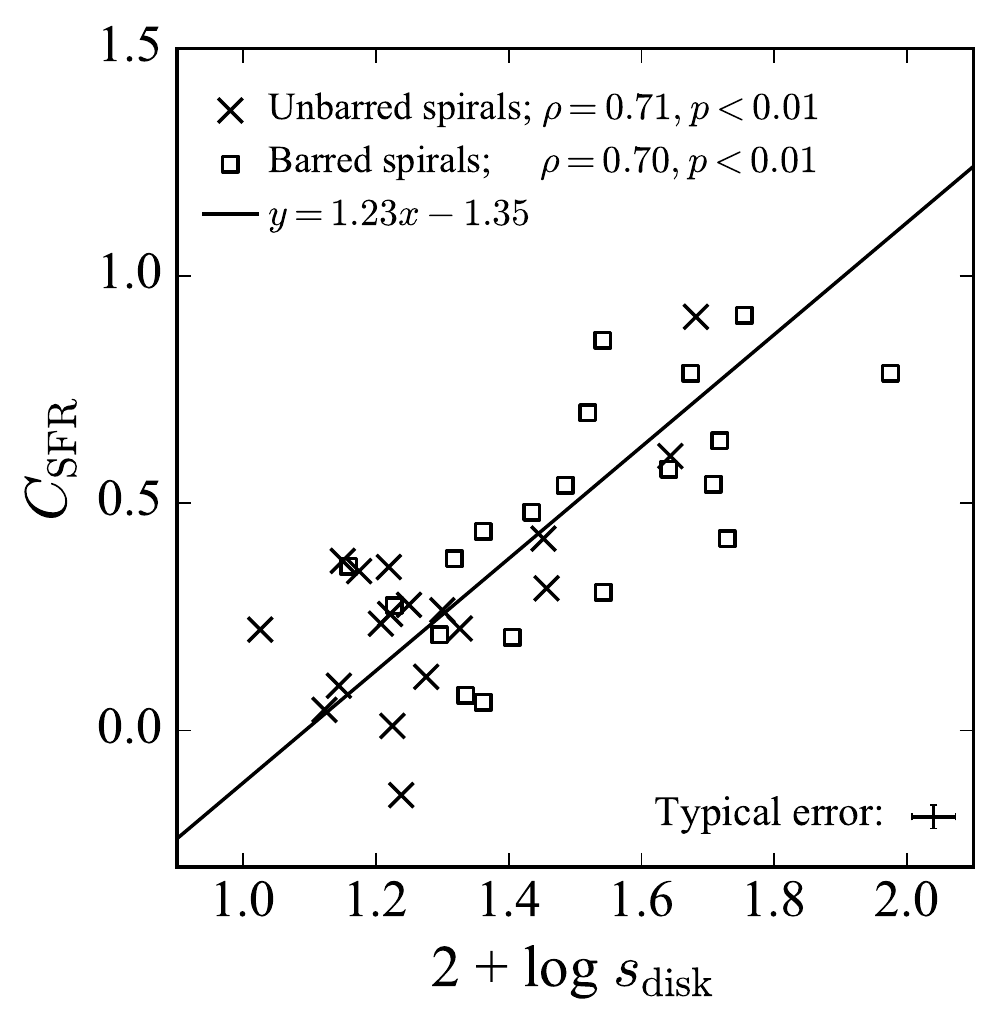}
        \caption{Dependence of central concentration of the SFR ($C_{\rm SFR}$) on the strength of non-axisymmetric disk structure ($2+\log s_{\rm disk}$). AGNs, LINERs and galaxies with a quenched center have been excluded. The Pearson correlation coefficients for unbarred and barred galaxies are given at the top of each panel. The formulas indicate the linear fits, derived using PCA and marked with solid lines.
        }
        \label{dsfr_str}
\end{figure}

Non-axisymmetric disk structures are essential to understand the galaxy secular evolution \citep{Kormendy2004}. As the two most generic structures, spirals and bars may play an important role in driving gas into the central regions. Here the gas provides the fuel for the central star formation, which leads to the growth of pseudo bulges \citep[e.g.,][]{Kormendy2004}. While the effects of bars has been extensively explored, the effects of spirals are less understood.

\subsection{Bar-driven gas inflow}

Theoretical models and simulations show that a stellar bar imposes a non-axisymmetric gravitational potential on the disk to drive the gas clouds flow down the dust lanes along the leading edge of the bar \citep{Athanassoula1992a, Athanassoula1992b, Regan1999, Sheth2000, Sheth2002, Regan2004, Combes2014, Fragkoudi2016, Tress2020, Sormani2020}. In a galaxy where there is a well-defined ILR introduced by the bar, the bar-driven inward flowing gas accumulates at the ILR at a distance of a few kiloparsecs from the center \citep{Athanassoula1984, Jenkins1994}; later the response of gas to forcing by the bar can give rise to a nuclear spiral or bar pattern in the gas, which can transport the gas to the proximity of the central black hole \citep{Englmaier2000, Englmaier2004}. Observations have confirmed the bar-driven gas transport \citep{Mundell1999, Combes2003, Zurita2004, Fathi2006, Jogee2006, Haan2009, Querejeta2016}. The rates of bar-driven gas inflow range from 0.01 to 50\,$M_{\odot}$/yr \citep{Regan1997, Haan2009, Querejeta2016}, leading to an accumulation of gas in the center \citep{Sheth2005}. The molecular gas concentrations in the central 1 kpc are systematically higher in barred spirals than in unbarred spirals \citep{Sakamoto1999b, Sheth2005, Kuno2007}. By comparing the molecular gas concentrations in barred and unbarred galaxies, it has been concluded that the bar is an efficient tool for driving gas to the galaxy center. Furthermore, this bar effect is more efficient in stronger bars \citep{Regan2004, Hopkins2011, KWT2012}. Consistently, galaxies hosting a stronger bar are found to have a more centrally concentrated gas distribution \citep{Kuno2007}.

These previous observational studies calculated molecular gas concentrations by assuming a constant $\alpha_{\rm CO}$ and they therefore actually measured CO luminosity concentrations. However, $\alpha_{\rm CO}$ depends on the conditions of gas clouds, such as, density, temperature, metallicity, and velocity dispersion \citep[e.g.,][]{Narayanan2012, Bolatto2013}.  Likely as a result of these effects, galaxy centers tend to have lower $\alpha_{\rm CO}$. The Milky Way has a $\alpha_{\rm CO}$ near its center that is 3-10 times lower than in the solar neighborhood \citep[e.g.,][]{Sodroski1995, Dahmen1998, Strong2004}. Observations have also found substantially lower $\alpha_{\rm CO}$ in nearby galaxy centers than the Milky Way disk value \citep[e.g.,][]{Sandstrom2013, Israel2020}. The $\alpha_{\rm CO}$ depression in the center reduces the molecular gas concentration compared to that derived using a constant $\alpha_{\rm CO}$. The degree of central depression in $\alpha_{\rm CO}$ varies from galaxy to galaxy, imposing different level of influence on molecular gas concentrations. To understand the effect of a variable $\alpha_{\rm CO}$ on molecular gas concentrations, we follow the suggestion in \cite{Bolatto2013} to derive the $\alpha_{\rm CO}(Z, \Sigma_{\star})$ according to the local metallicity and surface density (Sect.~\ref{Methods}). The resulting $\alpha_{\rm CO}(Z, \Sigma_{\star})$ is on average a factor of 4.5 lower at the center than the outer disk. We use the $\alpha_{\rm CO}(Z, \Sigma_{\star})$ to derive molecular gas concentrations ($C_{\rm mol}$). We find that the CO luminosity concentrations ($C_{\rm CO}$) tightly correlate with $C_{\rm mol}$ (Fig.~\ref{Clm}), but $C_{\rm mol}$ are systematically 0.22\,dex lower.  By accounting for the primary dependences of $\alpha_{\rm CO}$ on galaxy local properties, we confirm that high $C_{\rm mol}$ are more common in barred than unbarred galaxies and that barred galaxies with a stronger bar tend to have higher $C_{\rm mol}$ (Fig.~\ref{CStr}). A variable $\alpha_{\rm CO}$ therefore does not affect the conclusions drawn in \cite{Sakamoto1999b}, \cite{Sheth2005} and \cite{Kuno2007}.

Nevertheless, the KS test shows that $C_{\rm mol}$ in barred galaxies are not statistically significantly different from $C_{\rm mol}$ in unbarred galaxies. This is because there are some unbarred galaxies with relatively high $C_{\rm mol}$. To thoroughly explain why the high $C_{\rm mol}$ are more common in barred than in unbarred spirals, the spiral effect should be taken into account.

\subsection{Spiral-driven gas inflow}

Gas clouds are subjected to large-scale shocks upon passage across spiral arms that can trigger gravitational collapse and accelerate the formation of new stars \citep{Roberts1969, Kalnajs1972, Roberts1972}, which successfully explains the observed bright SF regions along the arms.  Hydrodynamical simulations of isolated disk with a prior input spiral density wave confirm the spiral shock on gas \citep{Kim2014, Kim3_2014, Kim2020, Baba2016, Pettitt2020}. The influence of spiral arms on gas clouds is implicit in the Silk-Elmegreen star formation law \citep{Elmegreen1997, Silk1997}, which proposes that multiple passages of gas clouds through the spiral gravitational potential favor the growth of gas clouds via cloud–cloud collision. In agreement with the compression of gas by shocks, \cite{Colombo2014} show that, in M51, the giant molecular clouds along the arms are brighter than those between the arms, and have a higher number density of more massive clouds, suggesting the arms of M51 promoting the cloud formation. Recently, \cite{Meidt2021} show that the logarithm of molecular arm-interarm contrasts, traced by CO emission observation, are higher than that of old stars, traced by 3.6\,$\mu$m observation, in a correlation steeper than linear even in the presence of weak or flocculent spiral arms.  Stronger spirals are expected to trigger stronger shocks \citep{Roberts1969, Kim2014}. Consistently, \cite{Seigar2002}, \cite{Kendall2015}, and \cite{Yu2021} have found that galaxies with strong spiral arms are found to have more intense global SFR or specific SFR than galaxies with weak arms.

Studies on spatially resolved star formation efficiency (SFE) present mixed results. The SFE across the density wave arms of M51 is reduced due to the streaming motion that introduces an additional stabilization effect on gas clouds \citep{Meidt2013}. The SFE enhancement along the arms of NGC~6946 is relatively well defined \citep{Rebolledo2012}, while the enhancement is absence in NGC~628 (\citealt{Kreckel2016}; see also \citealt{Foyle2010}).  Recently, \cite{Querejeta2021} show that spiral arms accumulate gas and star formation, without systematically increasing the SFE. Unlike the case of M51 \citep{Colombo2014}, the giant molecular cloud (GMC) properties change slightly but not significantly between arms and inter-arms in noninteracting galaxies \citep{Rosolowsky2021, Colombo2022}. Although the spiral arms of our Galaxy seem have little or no effect on the enhancement of the SFE \citep{Eden2015, Ragan2016, Urquhart2021}, evidence that supports the arms influencing the dense gas distribution or kinematics has been reported \citep{Sakai2015, Urquhart2021}. The complexity of star formation in the arms and interarm regions may be attributed to the turbulent and streaming motions induced when the shock increases the gas density, which prevents gas cloud collapse and reduces the SFE \citep{Meidt2013, Leroy2017, Yajima2019, Querejeta2019}. Spiral shocks are also essential for explaining the azimuthal offset between arm ridges of different components or for explaining the smaller winding angles of younger stellar arms \citep{Grosbol1998, Gittins2004, Egusa2009, MartinezGarcia2014, YuHo2018}.

The funneling of the gas toward the center is expected, as the shock causes the gas to lose angular momentum \citep{Kalnajs1972, Roberts1972, Lubow1986}. \cite{Lubow1986} included the back reaction of gas on stellar disk to study spiral shocks and found that parameters representing the solar neighborhood produced a gas accretion rate of $\sim$\,$(0.2$\textendash$0.4)\,M_{\odot}\,{\rm yr}^{-1}$, which is consistent with mass drift derived from chemical modeling in the Milky Way \citep{Lacey1985} and consistent with those in nearby galaxies derived based on gravitational torques \citep{Garcia-Burillo2009, Haan2009}.  The mass inflow is actually caused by a combination of three processes: angular momentum loss at spiral shocks ($\sim$\,50\% contribution), stellar spiral gravitational torque ($\sim$\,40\% contribution), and gaseous spiral gravitational torque ($\sim$\,10\% contribution; \citealt{Kim2014}).  By examining the nonlinear gas responses to an imposed stellar spiral potential with various properties in thin disks, \cite{Kim2014} show that the arms drive mass inflows at rates of $\sim$\,0.05\textendash3.0\,$M_{\odot}$\,yr$^{-1}$, with larger values corresponding to stronger and more slowly rotating arms. Although study of individual galaxies has shown that a few unbarred spiral galaxies have relatively high gas concentration \citep{Sheth2005, Regan2006}, the effect of spiral arms did not receive enough attention. By using a statistical large sample selected from the EDGE, we show that unbarred galaxies with stronger spiral arms tend to have more centrally concentrated molecular gas distribution, providing significant evidence for the spiral-driven transport of molecular gas to the central 2\,kpc of galaxies. We verify that it is not caused by the method adopted to identify the bars or induced by central bulges.

\subsection{Gas inflow driven by both spirals and bars}

By quantifying the strength of the disk (spiral+bar) structure, we find that the molecular gas concentration increases as the structure strengthens, regardless of whether they are unbarred or barred (Fig.~\ref{CStr}). Despite the higher gas concentrations and stronger structure found in barred galaxies, there is no significant difference between the correlations for barred and unbarred spirals, implying that the gas concentrations may be uniformly driven by shocks and torques produced by non-axisymmetric disk (spiral+bar) structures. The scatter in molecular gas concentration for given structure strength is $\sim$\,0.3\,dex. The scatter arises from several aspects. There are degeneracies between spiral strength and spiral pattern speed and degeneracies between bar strength and bar orbit morphologies. For a given spiral strength, spiral arms with a slower pattern speed trigger a stronger shock and a higher gas infall rate \citep{Kim2014}, and a higher gas concentration is expected. For a given bar strength, a bar could have different orbit morphologies, leading to various gas inflow rates and hence a variation in gas concentrations \citep{Regan2004}. The inflow of gas to the center may trigger a starburst at the center, which in turn consumes gas in a short timescale. This occurs periodically and is regulated by a balance between the inﬂow rate and central accumulation of the gas \citep{Krumholz2015}, perhaps leading to periodic changes in the molecular gas concentrations. The AGN-driven outflow, if present, may remove central gas content to reduce gas concentrations and cause some degree of variation in the $C_{\rm CO}$- or $C_{\rm mol}$-($2+\log s_{\rm disk}$) relation. The possible AGN feedback is discussed in Sect.~\ref{AGNH}.

We reinterpret the previous finding that barred galaxies tend to have higher molecular gas concentrations than unbarred galaxies \citep{Sakamoto1999b, Sheth2005, Kuno2007, Komugi2008}.  By assuming that the spiral- and bar-driven instabilities are the key drivers of gas inflow, the finding should be traced to the fact that barred spirals generally have stronger disk non-axisymmetric structure than unbarred spirals (Fig.~\ref{histSS}) and therefore have more gas transported to the center (Fig.~\ref{CStr}), leading to higher molecular gas concentrations.

\subsection{Gas concentrations in AGN hosts}\label{AGNH}

AGNs may transfer radiative or mechanical energy to the surrounding gas and blow cold gas out of the galaxy \citep[e.g.,][]{Silk1998, Croton2006, Zinger2020}. However, studies of low redshift AGNs have shown that, compared to inactive galaxies of similar mass and morphology, the AGN hosts have similar amounts of atomic \citep{Ho2008, Fabello2011, Ellison2019}, molecular \citep{Maiolino1997, Saintonge2017, Shangguan2020, Yesuf2020, Koss2021}, and total gas content \citep{Vito2014,Shangguan2018, Shangguan2019}. These results suggest that low redshift AGN feedback is inefficient at removing gas, or, alternatively, AGN feedback effects may be limited to the galaxy central regions. Outflows within a few kiloparsecs from the center have been identified in AGN hosts \citep{Karouzos2016, Kang2018, Fluetsch2019}, although the contribution of AGN outflow to feedback is unclear \citep{Woo2017}. \cite{Ellison2021} show that the gas fractions of central AGN regions are lower than those in SF regions and they suggest that the expelling of gas by AGN feedback may happen on central sub-galactic kiloparsec scales instead of global scale. If this is true, AGN hosts should have notably lower gas concentrations compared to inactive galaxies of similar mass or morphology.

We test the effect of AGNs on the radial distribution of molecular gas using the $C_{\rm CO}$- and $C_{\rm mol}$-($2+\log s_{\rm disk}$) relations. In Fig.~\ref{SFSey} we plot the $C_{\rm CO}$- and $C_{\rm mol}$-($2+\log s_{\rm disk}$) relations separately for 30 pure SF, 15 composite, and 4 Seyfert galaxies. The solid lines in each panel all mark the best-fit linear function of the relation for pure SF galaxies. The $C_{\rm CO}$- and $C_{\rm mol}$-($2+\log s_{\rm disk}$) relations for pure SF galaxies are consistent within uncertainty with those constructed using the full sample in Fig.~\ref{CStr}.  The composite galaxies have similar ranges of $C_{\rm CO}$ and $C_{\rm mol}$, and do not have significantly lower $C_{\rm CO}$ or $C_{\rm mol}$ compared to that in SF galaxies. They lie around the SF best-fit functions.

Among our 4 Seyferts, there is one type~1 AGN (UGC~3973) and three type~2 AGNs (NGC~2410, NGC~2639, NGC~6394), as identified by \cite{Lacerda2020}. NGC~2410 and NGC~2639 are strong AGNs \citep{Kalinova2021}. There are one Seyferts lie above the SF best-fit function, and three lie below it. \cite{Ellison2021} show that NGC~2639 has significantly lower local gas fraction in the AGN region than in SF regions. Consistent with their results, this galaxy (bottom-left corner on the diagrams for Seyferts) has the lowest gas concentrations: $C_{\rm CO}$\,$=$\,$-0.22$ and $C_{\rm mol}$\,$=$\,$-0.52$. However, it also has almost the weakest structure strength: $2+\log s_{\rm disk}$\,$=$\,$0.83$. The reduction of the central gas content in this galaxy may be caused by its AGN feedback \citep{Ellison2021}.  However, because NGC~2639 lacks a bar and its spiral arms almost vanish, the inflow of gas driven by spiral arms and bars ceases. The low molecular gas concentration observed in NGC~2639 can therefore be explained by the lack of disk instabilities that transport gas inward. Taking the best-fit function for the pure SF galaxies as a reference, the $C_{\rm CO}$ or $C_{\rm mol}$ in Seyferts tend to be lower but not significant.

We calculate the difference ($\Delta C_{\rm CO}$) between the measured $C_{\rm CO}$ and the best-fit function constructed using pure SF galaxies at the given measured $2+\log s_{\rm disk}$, and the difference for $C_{\rm mol}$ ($\Delta C_{\rm mol}$) in the same way. Figure~\ref{del_C} presents the histogram of $\Delta C_{\rm CO}$ and $\Delta C_{\rm mol}$ for pure SF, composite, and Seyfert galaxies. The KS test between $\Delta C_{\rm CO}$ in pure SF and composite galaxies yields a $p_{\rm KS}$\,$=$\,$0.94$, and the same for $\Delta C_{\rm mol}$ yields a $p_{\rm KS}$\,$=$\,$0.63$. This suggests that we cannot reject the null hypothesis that the $\Delta C_{\rm CO}$ in pure SF and composite galaxies come from the same parent distribution. Likewise, the KS test between $\Delta C_{\rm CO}$ in pure SF and Seyfert galaxies yields a $p_{\rm KS}$\,$=$\,$0.34$, and the same for $\Delta C_{\rm mol}$ yields a $p_{\rm KS}$\,$=$\,$0.12$. This suggests that we cannot reject the null hypothesis that the $\Delta C_{\rm CO}$ in pure SF and Seyfert galaxies come from the same parent distribution. It is important to note that the size of the Seyferts subsample is small, which makes our results less reliable. Future analyses of more Seyferts may strengthen our findings.

We therefore do not reveal a significant drop in gas concentrations of Seyferts compared to non-AGN galaxies of similar strength in our sample. If the shocks and torques produced by non-axisymmetric structures such as spiral arms and bars are key factors to regulate the molecular gas distribution, our results suggest that the AGN feedback in Seyferts do not have a notable effect on the radial distribution of molecular gas on central $\sim$\,2\,kpc scales. Still, the AGN feedback of Seyferts in our sample may not be powerful enough or the feedback may work on smaller scale, study of which requires higher resolution observation.

\subsection{Molecular gas concentration and star formation rate concentration}

Star formation on global and sub-galactic scales couples directly with the molecular gas \citep[e.g.,][]{Kennicutt1989, Bigiel2008, Leroy2008, Leroy2013, Schruba2011, Bolatto2017}. A centrally concentrated distribution of SFR is therefore expected for a galaxy with centrally concentrated molecular gas. We focus on the subsample of  38 non-AGN galaxies with robust SFR measurements (Sect.~\ref{Sec_Csfr}). In this subsample, AGNs, ambiguous galaxies, and galaxies with central EW$_{\rm H\alpha}$\,$<$\,6\,$\AA$ are excluded. By defining $C_{\rm CO}$, $C_{\rm mol}$, and $C_{\rm SFR}$ over the same radial extent, we find that both $C_{\rm CO}$ and $C_{\rm mol}$ tightly correlate with $C_{\rm SFR}$, regardless of whether the galaxies are barred or unbarred ($\rho$\,$=$\,0.84\textendash0.91), as illustrated in Fig.~\ref{dsfr_cog}. The variable $\alpha_{\rm CO}(Z,\Sigma_{\star})$ does not influence the relationship strength between molecular gas concentrations and SFR concentrations.

\cite{Chown2019} find a loose correlation between gas concentrations and central upturn of EW$_{\rm H\alpha}$ profile, a measure of centrally enhanced SFR, in barred galaxies ($\rho$\,$=$\,0.64), and cannot find a similar one in unbarred galaxies ($\rho$\,$=$\,0.09). The failure to find a tight correlation in \cite{Chown2019} is due to the following two reasons. First, they adopted different radial scales to calculate the two quantities. Their gas concentration was defined over a mean radial extent $R_{\rm 50\%, CO}$\,$\approx$\,4\,kpc, while they measured the upturn of EW$_{\rm H\alpha}$ at $R$\,$=$\,$0$.  Second, determining by eye the turning point of the EW$_{\rm H\alpha}$ profile to calculate its central upturn introduces uncertainty. For instance, the EW$_{\rm H\alpha}$ profile of NGC~7819 has at least two turn-up and one turnover, and the turn-up at $R$\,$\approx$\,$7\farcs7$ is chosen by \cite{Chown2019}. However, this chosen turning is the weakest one among others. We use the change point analysis described in \cite{Watkins2019} to estimate the amplitude of the turning. Instead, if the strongest turning at $R$\,$\approx$\,$16\arcsec$ is used, NGC~7819 has a strong EW$_{\rm H\alpha}$ central upturn $\sim$\,1 rather than 0.3 reported by \citealt{Chown2019}. The EW$_{\rm H\alpha}$ central upturn of $\sim$\,1 is consistent with our result that NGC~7819 has a highly centrally concentrated SFR distribution.

By performing PCA, we find the best-fit straight lines for the $C_{\rm CO}$- and $C_{\rm mol}$-$C_{\rm SFR}$ relations (Fig.~\ref{dsfr_cog}) and the two relations have a small scatter of 0.04\,dex. Together with the $C_{\rm CO}$- and $C_{\rm mol}$-$(2+\log s_{\rm disk})$ relations (Fig.~\ref{CStr}), it is expected to see a connection between the central concentration of SFR and the strength of non-axisymmetric disk structure.

\subsection{Central concentrations of SFR driven by spiral arms and bars}

We probe the dependence of $C_{\rm SFR}$ on the disk structure strength in Fig.~\ref{dsfr_str}. First, similar to $C_{\rm CO}$ and $C_{\rm mol}$ shown in Fig.~\ref{CStr}, $C_{\rm SFR}$ for unbarred galaxies are mainly moderate to weak, whereas those for barred galaxies span a wide range and can reach particularly high values. Highly centrally concentrated SFRs are more common in barred than in unbarred galaxies. Second, the central concentration of SFR increases as the disk (spiral+bar) structure strengthens in both barred and unbarred galaxies. The relation for unbarred and barred galaxies has a correlation coefficient $\rho$\,$=$\,$0.71$ ($p$\,$<$\,$0.01$) and $\rho$\,$=$\,$0.70$ ($p$\,$<$\,$0.01$), respectively. We perform PCA to obtain the best-fit line:
\begin{equation}\label{Csfr_s}
  C_{\rm SFR} = (1.23\pm 0.20) \times (2+\log s_{\rm disk}) - (1.35\pm 0.27). 
\end{equation}
Uncertainties are obtained through bootstrapping with replacement. The scatter in $C_{\rm SFR}$ for a given structure strength is 0.18\,dex. Bar identification based on isophotal analysis (Appendix~\ref{ISO}) or 2D decomposition from \cite{JMA2017} yields consistent results, although the correlation coefficients vary slightly (row [3] in Table~\ref{tid}). Morphological quenching proposes that the bulges stabilizes the central gaseous disk to prevent the formation of new stars \citep{Martig2009}. We test the effect of morphological quenching using partial correlation coefficients. The residual dependence of $C_{\rm SFR}$ on $2+\log s_{\rm disk}$ after removing the effect of $B/T$ yields $\rho^{\prime}$\,$=$\,$0.62$ ($p$\,$<$\,$0.01$) in unbarred galaxies and $\rho^{\prime}$\,$=$\,$0.70$ ($p$\,$<$\,$0.01$) in barred galaxies (row [3] in Table~\ref{trho}). The reported correlations are not introduced by bulges or no obvious signature of morphological quenching is detected, perhaps because galaxies with a quenched center have been excluded by requiring central EW$_{\rm H\alpha}$\,$\geq$\,6\,$\AA$ (Sect.~\ref{Sec_Csfr}). After removing the effect of $C_{\rm CO}$ or $C_{\rm mol}$, no residual dependence of $C_{\rm SFR}$ on $2+\log s_{\rm disk}$ is found (rows [4] and [5] in Table~\ref{trho}). This suggests that the disk instabilities drive gas inflow to the centers, and the increased gas surface density raises central SFR surface density without significantly changing the central SFE. It is consistent with the result that the SFE derived based on molecular gas does not significantly change with radius \citep{Leroy2008, Muraoka2019}.

Our reported relationships are in agreement with previous findings.
The pioneering works by \cite{Martinet1997} and \cite{Aguerri1999} found a tight correlation between bar strength and global SF activity in noninteracting barred galaxies. Further, \cite{Zhou2015} and \cite{Lin2017} showed that star formation activity in the center is enhanced in galaxies with stronger bars (see also \citealt{James2009}). Strong bars may have suppressed central star formation \citep{Wang2012, Consolandi2017, Simon2020, Wang2020}, especially in those massive gas-poor galaxies \citep{Simon2020} with particularly weak spiral arms \citep{Wang2020}. However, these gas-poor galaxies with star formation that have been quenched are not of interest to the present work because the effects of gas inflow driven by disk instabilities have been removed. Consistent with our results, \cite{Yu2022} used a ratio of central fiber specific SFR acquired from the MPA-JHU catalog \citep{Brinchmann2004} to global specific SFR derived from \cite{Salim2018} as a measure of central enhancement of star formation, and found enhanced central SFR in galaxies with strong spiral arms.

We show that barred and unbarred galaxies with strong non-axisymmetric disk (spiral+bar) structures tend to have highly concentrated distribution of both molecular gas and SFR, providing evidence for bar-driven and, especially, spiral-driven gas transport to the central region. The accumulation of gas in the center boosts central SFR. It is consistent with the concept of secular evolution, which describes the slow rearrangement of energy and mass resulting from interactions facilitated by non-axisymmetric galaxy structures \citep{Combes1981,  Kormendy1982, Pfenniger1990, Sellwood1993, Kormendy2004}. \cite{Yu2022} also shed light on this perspective. They found that the stronger the arm structure, the more enhanced central SFR, the higher the fraction of galaxies hosting a pseudo bulge. The gas inflow triggered by disk instabilities drives secular growth of the central pseudo bulges, and the spirals and bars therefore play an essential role in the galaxy secular evolution.

\section{Summary and conclusions}\label{conclusions}

Models and simulations have shown that the instabilities induced by spirals and bars  are efficient mechanisms for transporting gas to the central regions, with the inflow rates depending on the perturbation strength \citep[e.g.,][]{Athanassoula1992a, Athanassoula1992b, Hopkins2011, Kim2014}. Observations supporting bar-driven gas inflow have shown that barred galaxies tend to have higher molecular gas concentrations than unbarred galaxies \citep{Sakamoto1999b, Sheth2005, Kuno2007, Komugi2008} and that stronger bars are associated with higher molecular gas concentrations \citep{Kuno2007}. These previous studies used a constant CO-to-H$_2$ conversion factor ($\alpha_{\rm CO}$) to derive molecular gas concentrations. However, the $\alpha_{\rm CO}$ depends on the physical properties of the environment where the gas clouds are embedded \citep[e.g.,][]{Narayanan2012, Bolatto2013}. It is not clear how a parameter-dependent $\alpha_{\rm CO}$ influences the calculation of molecular gas concentrations and the conclusions drawn in previous studies. Although spiral-driven gas inflow is predicted \citep{Kalnajs1972, Roberts1972, Lubow1986, Kim2014}, less is known about the spiral effect on the radial distribution of molecular gas.

We used a sample of 57 disk galaxies selected from the EDGE-CALIFA survey to investigate the connection between molecular gas concentrations and non-axisymmetric disk (spiral+bar) structures. The structure strength is defined as the logarithm of the average relative amplitude of spiral arms and bars, and it therefore traces average spiral strength for unbarred galaxies but average spiral and bar strength for barred galaxies. Molecular gas mass surface density is derived using a $\alpha_{\rm CO}$ that decreases with higher metallicity and higher stellar surface density, following the prescription in \cite{Bolatto2013}.The $\alpha_{\rm CO}$ at the center is on average a factor of 4.5 lower than in the outer disk. The central concentrations of CO luminosity ($C_{\rm CO}$) and molecular gas ($C_{\rm mol}$) and the SFR ($C_{\rm SFR}$) are defined, respectively, as the logarithmic ratio of the average central ($<$\,0.2\,$R_e$) CO intensity, the molecular mass surface density, and the SFR surface density to their disk-averaged ($<$\,$R_e$) surface densities. The 0.2\,$R_e$ on average corresponds to $\sim$\,1.1\,kpc in our sample, and the $C_{\rm CO}$, $C_{\rm mol}$, and $C_{\rm SFR}$ therefore measure concentrations in the central $\sim$\,2\,kpc. Our results are independent of methods used to identify bars and are not caused by bulges. The main findings are as follows.

\begin{enumerate}
  \item By construction, $C_{\rm mol}$ and $C_{\rm CO}$ are tightly correlated, with $C_{\rm mol}$ 0.22\,dex lower on average due to the lower central region value of $\alpha_{\rm CO}$. The tight correlation between $C_{\rm mol}$ and $C_{\rm CO}$ implies that several conclusions obtained using a constant $\alpha_{\rm CO}$ in previous studies still stand, although they overestimate central molecular gas concentrations. We therefore confirm the previous finding that high $C_{\rm mol}$ values are more common in barred galaxies. However, a few unbarred galaxies with strong spiral arms can also have high $C_{\rm mol}$.
  \item There is a good correlation between $C_{\rm mol}$ and the strength of non-axisymmetric structure, which can be due to a bar, spiral arms, or both. It supports the idea that central gas concentrations are due to bar-driven and spiral-driven gas transport.
   \item Despite the small subsample size, the $C_{\rm mol}$ of four Seyferts are not significantly reduced compared to inactive galaxies of similar disk structure. This suggests that the AGN feedback in Seyferts may not notably affect the molecular gas distribution in the central $\sim$\,2\,kpc. 
   \item $C_{\rm mol}$ tightly correlates with $C_{\rm SFR}$. Galaxies with stronger disk (spiral+bar) structure tend to have higher $C_{\rm SFR}$, regardless of whether they are barred or unbarred. After removing the effect of $C_{\rm mol}$, no residual dependence of $C_{\rm SFR}$ on structure strength is found, suggesting that spiral arms and bars transport gas to the centers but have no additional significant effect on further raising the central SFE.

\end{enumerate}

Our results provide significant evidence for bar-driven and, especially, spiral-driven gas transport to the central regions of galaxies. The accumulation of gas in the center increases the central SFR, which likely facilitates the buildup of central disky pseudo bulges. Both spiral arms and bars therefore play an essential role in the secular evolution of disk galaxies.

\begin{acknowledgements}
We are grateful to the anonymous referee for helpful feedback.
SYY is indebted to Karl Menten for his great support during the pandemic. 
SYY acknowledges support by the Alexander von Humboldt Foundation.
RCL acknowledges support provided by a National Science Foundation (NSF) Astronomy and Astrophysics Postdoctoral Fellowship under award AST-2102625.
LCH was supported by the National Science Foundation of China (11721303, 11991052, 12011540375) and the China Manned Space Project (CMS-CSST-2021-A04 and CMS-CSST-2021-A06).  We beneﬁted from discussions with Adam K. Leroy.
Support for CARMA construction was derived from the Gordon and Betty Moore Foundation, the Eileen and Kenneth Norris Foundation, the Caltech Associates, the states of California, Illinois, and Maryland, and the NSF. Funding for CARMA development and operations were supported by NSF and the CARMA partner universities.
\end{acknowledgements}

\bibliographystyle{aa}

% \bibliography{syu.bib}

%############################################################

\begin{appendix}

\section{Table of derived quantities}
Table~\ref{tdata} presents the quantities derived in Sect.~\ref{Methods}.

\longtab[1]{
\renewcommand\arraystretch{1.2}
\begin{longtable}{ccccccccccc}
\caption{\label{tdata} Table of derived quantities}\\
\hline
\hline
Name & Bar & $\epsilon_{\rm disk}$ & PA$_{\rm disk}$ &  $2+\log s_{\rm disk}$ & $C_{\rm CO}$ & $C_{\rm mol}$  & $C_{\rm SFR}$ & Nuclear activity \\
(1)&(2)&(3)&(4)&(5)&(6)&(7)&(8)&(9)\\
\hline
\endfirsthead
\caption{continued}\\
\hline
Name & Bar & $\epsilon_{\rm disk}$ & PA$_{\rm disk}$ &  $2+\log s_{\rm disk}$ & $C_{\rm CO}$ & $C_{\rm mol}$  & $C_{\rm SFR}$ & Nuclear activity \\
\hline
\endhead
\hline
\endfoot
\hline
\endlastfoot
IC~944 & N & 0.66 & 105 & $1.29\pm 0.01$ & $0.39\pm 0.05$ & $0.23\pm 0.04$ & $\dots$ & Composite \\
IC~1199 & B & 0.63 & 158 & $1.16\pm 0.03$ & $0.35\pm 0.03$ & $0.21\pm 0.02$ & $0.36\pm 0.03$ & Pure SF \\
IC~1683 & B & 0.44 & 13 & $1.71\pm 0.04$ & $0.49\pm 0.03$ & $0.33\pm 0.03$ & $0.54\pm 0.03$ & Pure SF \\
IC~4566 & B & 0.31 & 152 & $1.52\pm 0.06$ & $-0.06\pm 0.03$ & $-0.40\pm 0.04$ & $\dots$ & Ambiguous \\
NGC~447 & B & 0.06 & 94 & $1.69\pm 0.03$ & $1.01\pm 0.02$ & $0.88\pm 0.02$ & $\dots$ & Ambiguous \\
NGC~477 & B & 0.47 & 124 & $1.73\pm 0.06$ & $0.45\pm 0.02$ & $0.18\pm 0.03$ & $0.42\pm 0.04$ & Pure SF \\
NGC~496 & N & 0.44 & 32 & $1.22\pm 0.03$ & $0.48\pm 0.03$ & $0.37\pm 0.03$ & $0.36\pm 0.03$ & Pure SF \\
NGC~551 & B & 0.57 & 136 & $1.30\pm 0.04$ & $0.03\pm 0.03$ & $-0.16\pm 0.04$ & $\dots$ & Composite \\
NGC~2253 & B & 0.23 & 130 & $1.36\pm 0.03$ & $0.54\pm 0.01$ & $0.31\pm 0.01$ & $0.44\pm 0.02$ & Pure SF \\
NGC~2347 & B & 0.38 & 4 & $1.23\pm 0.03$ & $0.29\pm 0.02$ & $0.05\pm 0.01$ & $0.27\pm 0.03$ & Composite \\
NGC~2410 & B & 0.71 & 32 & $1.18\pm 0.04$ & $0.31\pm 0.04$ & $0.03\pm 0.02$ & $\dots$ & Seyfert \\
NGC~2487 & B & 0.05 & 14 & $1.57\pm 0.03$ & $0.91\pm 0.01$ & $0.58\pm 0.01$ & $\dots$ & Ambiguous \\
NGC~2639 & N & 0.47 & 133 & $0.83\pm 0.04$ & $-0.22\pm 0.03$ & $-0.52\pm 0.03$ & $\dots$ & Seyfert \\
NGC~2730 & B & 0.30 & 84 & $1.54\pm 0.04$ & $0.37\pm 0.02$ & $0.22\pm 0.02$ & $0.30\pm 0.01$ & Pure SF \\
NGC~2906 & N & 0.43 & 82 & $1.28\pm 0.03$ & $0.35\pm 0.02$ & $0.10\pm 0.02$ & $\dots$ & Composite \\
NGC~3381 & B & 0.16 & 52 & $1.67\pm 0.01$ & $0.72\pm 0.02$ & $0.53\pm 0.02$ & $0.79\pm 0.01$ & Pure SF \\
NGC~3687 & B & 0.11 & 142 & $1.23\pm 0.03$ & $0.50\pm 0.04$ & $0.09\pm 0.05$ & $\dots$ & Ambiguous \\
NGC~3811 & B & 0.24 & 173 & $1.54\pm 0.04$ & $0.85\pm 0.02$ & $0.61\pm 0.01$ & $0.86\pm 0.01$ & Pure SF \\
NGC~3815 & N & 0.56 & 70 & $1.24\pm 0.06$ & $0.03\pm 0.03$ & $-0.18\pm 0.02$ & $-0.14\pm 0.03$ & Pure SF \\
NGC~3994 & B & 0.48 & 8 & $1.34\pm 0.02$ & $0.10\pm 0.02$ & $-0.04\pm 0.01$ & $0.08\pm 0.01$ & Composite \\
NGC~4047 & N & 0.21 & 97 & $1.03\pm 0.03$ & $0.29\pm 0.02$ & $0.11\pm 0.02$ & $0.22\pm 0.02$ & Pure SF \\
NGC~4185 & B & 0.33 & 164 & $1.13\pm 0.04$ & $0.02\pm 0.04$ & $-0.23\pm 0.04$ & $\dots$ & Composite \\
NGC~4210 & B & 0.25 & 97 & $1.37\pm 0.04$ & $0.03\pm 0.03$ & $-0.25\pm 0.03$ & $\dots$ & Ambiguous \\
NGC~4470 & N & 0.33 & 178 & $1.14\pm 0.04$ & $0.07\pm 0.04$ & $-0.05\pm 0.03$ & $0.10\pm 0.02$ & Pure SF \\
NGC~4711 & N & 0.48 & 40 & $1.22\pm 0.03$ & $0.20\pm 0.03$ & $0.00\pm 0.03$ & $0.01\pm 0.03$ & Composite \\
NGC~4961 & B & 0.33 & 102 & $1.36\pm 0.03$ & $0.23\pm 0.03$ & $0.08\pm 0.03$ & $0.06\pm 0.02$ & Pure SF \\
NGC~5000 & B & 0.05 & 90 & $1.75\pm 0.03$ & $0.91\pm 0.02$ & $0.79\pm 0.01$ & $0.91\pm 0.02$ & Composite \\
NGC~5016 & N & 0.25 & 54 & $1.17\pm 0.05$ & $0.26\pm 0.02$ & $0.05\pm 0.01$ & $0.35\pm 0.02$ & Pure SF \\
NGC~5056 & B & 0.45 & 0 & $1.41\pm 0.04$ & $0.47\pm 0.03$ & $0.24\pm 0.02$ & $0.20\pm 0.02$ & Composite \\
NGC~5205 & B & 0.44 & 164 & $1.50\pm 0.03$ & $0.17\pm 0.04$ & $-0.18\pm 0.06$ & $\dots$ & Composite \\
NGC~5406 & B & 0.28 & 117 & $1.55\pm 0.04$ & $-0.04\pm 0.05$ & $-0.43\pm 0.06$ & $\dots$ & Ambiguous \\
NGC~5480 & N & 0.19 & 26 & $1.64\pm 0.04$ & $0.50\pm 0.02$ & $0.31\pm 0.02$ & $0.60\pm 0.01$ & Pure SF \\
NGC~5520 & N & 0.50 & 64 & $1.15\pm 0.03$ & $0.29\pm 0.03$ & $0.04\pm 0.02$ & $0.37\pm 0.03$ & Pure SF \\
NGC~5633 & N & 0.33 & 14 & $1.12\pm 0.03$ & $0.12\pm 0.01$ & $-0.02\pm 0.01$ & $0.05\pm 0.01$ & Pure SF \\
NGC~5657 & B & 0.62 & 165 & $1.64\pm 0.04$ & $0.42\pm 0.04$ & $0.31\pm 0.02$ & $0.57\pm 0.04$ & Pure SF \\
NGC~5732 & N & 0.46 & 42 & $1.22\pm 0.03$ & $0.22\pm 0.04$ & $0.11\pm 0.04$ & $0.26\pm 0.02$ & Pure SF \\
NGC~5784 & N & 0.18 & 24 & $1.23\pm 0.05$ & $0.33\pm 0.02$ & $0.08\pm 0.02$ & $\dots$ & Ambiguous \\
NGC~5947 & B & 0.18 & 65 & $1.48\pm 0.03$ & $0.29\pm 0.02$ & $-0.07\pm 0.02$ & $0.54\pm 0.01$ & Composite \\
NGC~5980 & N & 0.63 & 12 & $1.21\pm 0.03$ & $0.28\pm 0.04$ & $0.11\pm 0.04$ & $0.23\pm 0.03$ & Pure SF \\
NGC~6004 & B & 0.21 & 94 & $1.43\pm 0.03$ & $0.69\pm 0.02$ & $0.38\pm 0.02$ & $0.48\pm 0.01$ & Composite \\
NGC~6060 & N & 0.57 & 100 & $1.25\pm 0.03$ & $0.48\pm 0.03$ & $0.18\pm 0.03$ & $0.28\pm 0.05$ & Pure SF \\
NGC~6155 & N & 0.31 & 142 & $1.28\pm 0.04$ & $0.23\pm 0.02$ & $0.05\pm 0.01$ & $0.12\pm 0.01$ & Pure SF \\
NGC~6186 & B & 0.18 & 37 & $1.98\pm 0.03$ & $0.70\pm 0.03$ & $0.56\pm 0.03$ & $0.79\pm 0.03$ & Pure SF \\
NGC~6301 & N & 0.41 & 108 & $1.26\pm 0.03$ & $0.07\pm 0.03$ & $-0.15\pm 0.03$ & $\dots$ & Pure SF \\
NGC~6394 & B & 0.62 & 40 & $1.58\pm 0.04$ & $0.17\pm 0.04$ & $0.04\pm 0.04$ & $\dots$ & Seyfert \\
NGC~6478 & N & 0.61 & 33 & $1.30\pm 0.04$ & $0.25\pm 0.04$ & $0.02\pm 0.03$ & $0.26\pm 0.04$ & Composite \\
NGC~7738 & B & 0.33 & 68 & $2.08\pm 0.04$ & $0.78\pm 0.04$ & $0.57\pm 0.04$ & $\dots$ & Ambiguous \\
NGC~7819 & N & 0.34 & 107 & $1.68\pm 0.03$ & $0.85\pm 0.03$ & $0.70\pm 0.03$ & $0.91\pm 0.03$ & Pure SF \\
UGC~3253 & B & 0.45 & 85 & $1.52\pm 0.03$ & $0.66\pm 0.03$ & $0.49\pm 0.02$ & $0.70\pm 0.04$ & Composite \\
UGC~3973 & B & 0.20 & 159 & $1.88\pm 0.03$ & $0.50\pm 0.02$ & $0.13\pm 0.01$ & $\dots$ & Seyfert \\
UGC~4132 & B & 0.70 & 28 & $1.32\pm 0.01$ & $0.36\pm 0.05$ & $0.26\pm 0.04$ & $0.38\pm 0.04$ & Pure SF \\
UGC~4461 & N & 0.74 & 43 & $1.45\pm 0.03$ & $0.49\pm 0.06$ & $0.35\pm 0.05$ & $0.42\pm 0.04$ & Pure SF \\
UGC~5108 & B & 0.59 & 139 & $1.72\pm 0.04$ & $0.47\pm 0.04$ & $0.27\pm 0.03$ & $0.64\pm 0.08$ & Composite \\
UGC~5359 & B & 0.63 & 94 & $1.41\pm 0.05$ & $0.04\pm 0.04$ & $-0.23\pm 0.03$ & $\dots$ & Pure SF \\
UGC~9067 & B & 0.53 & 12 & $1.30\pm 0.04$ & $0.20\pm 0.03$ & $0.06\pm 0.02$ & $0.21\pm 0.02$ & Pure SF \\
UGC~9476 & N & 0.34 & 133 & $1.46\pm 0.04$ & $0.41\pm 0.02$ & $0.19\pm 0.02$ & $0.31\pm 0.01$ & Pure SF \\
UGC~9542 & N & 0.71 & 33 & $1.33\pm 0.03$ & $0.34\pm 0.06$ & $0.14\pm 0.05$ & $0.22\pm 0.04$ & Pure SF \\
\end{longtable}
\tablefoot{Column (1): galaxy name. 
Column (2): bar type. ``N'' denotes unbarred galaxies, while ``B'' denotes barred galaxies. The bar classification is from \cite{Walcher2014}.
Column (3): ellipticity of the galaxy. 
Column (4): position angle of the galaxy in degrees. 
Column (5): strengths of non-axisymmetric disk structure and the uncertainty.
Column (6): central concentrations of CO luminosity and the uncertainty.
Column (7): central concentrations of molecular gas and the uncertainty.
Column (8): central concentrations of SFR and the uncertainty.
Column (9): types of nuclear activity. }
}

%----------------------------------------------------------------------------
\section{Bar identification based on isophotal analysis} \label{ISO}

%%--------------------------------------------------------------------

\begin{figure}
        \centering
        \includegraphics[width=0.5\textwidth]{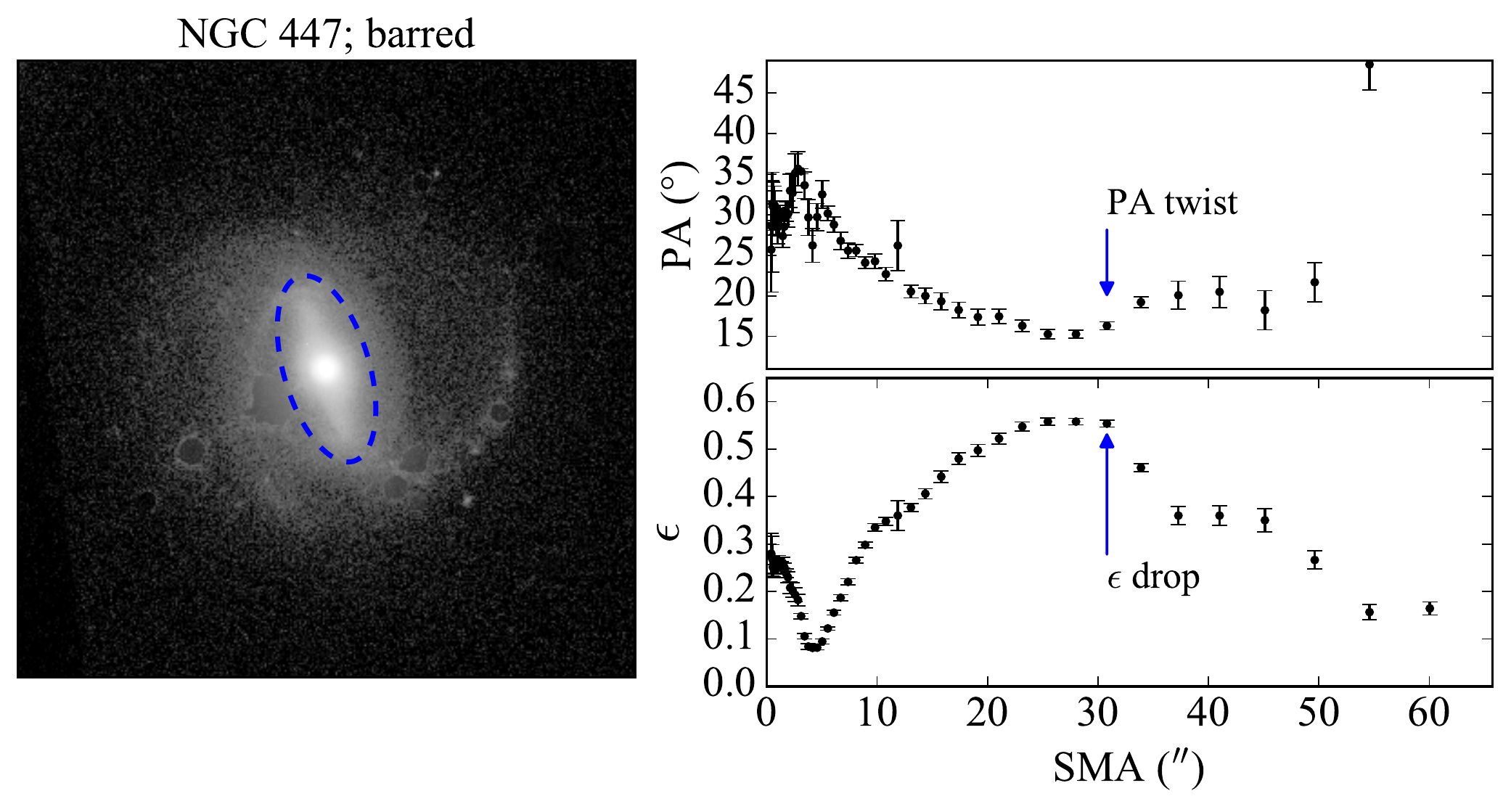}
        \caption{Illustration of how to identify a bar using profiles of $\epsilon$ and PA. The ellipse in the left panel denotes the bar with measured $\epsilon_{\rm bar}$, PA$_{\rm bar}$ and SMA$_{\rm bar}$. The arrows in the bottom-right and top-right panels denotes the $\epsilon$ drop and the PA twist caused by the bar, respectively.  
        }
        \label{app_exmp}
\end{figure}

We perform an additional bar identification based on isophotes. It has been found that the $\epsilon$ profile tends to increase with semimajor axis (SMA) and the PA profile tend to keep constant within the region dominated by a bar, beyond which the $\epsilon$ would drop and PA would twist \citep{Athanassoula2002, Laine2002, Erwin2003, Menendez-Delmestre2007, Aguerri2009, Li2011}. We thus search for drops in the $\epsilon$ profile and twists in the PA profile as bar candidates. We inspect the images to ensure that dust lanes, star formation, rings, or spiral arms are not responsible for these features in the $\epsilon$ and PA profiles. A candidate bar is identified as a true bar only if there is a bar-like structure in the {\it r}-band image consistent with its $\epsilon$, PA, and SMA. We give preference to the bar candidate with maximum $\epsilon$. Figure~\ref{app_exmp} illustrates the bar identification for NGC~447. The ellipse in the left panel denotes the bar with measured $\epsilon_{\rm bar}$, PA$_{\rm bar}$, and SMA$_{\rm bar}$; the arrow in the middle and right panel denotes the $\epsilon$ drop and PA twist caused by the bar. Then thePA$_{\rm bar}$ and SMA$_{\rm bar}$ are used to calculate the intrinsic bar radius:
\begin{equation}
  R_{\rm bar}={\rm SMA_{bar}}\sqrt{(  \cos\,\Delta {\rm PA} )^2+\left( \frac{\sin\,\Delta {\rm PA} }{1-\epsilon_{\rm disk}}  \right)^2},
\end{equation}
\noindent
 where $\Delta {\rm PA=PA_{disk}-PA_{bar}}$. The vertical dashed line in the top-right panel of Fig.~\ref{exmp} denotes the $R_{\rm bar}$ for NGC~447.  The isophotal analysis yields a bar fraction of only 42\% and it may have missed some weak bars. The isophote-based bar identification is used to understand the uncertainty in our results introduced by using different methods to identify bars (Sect.~\ref{barid}).

\begin{figure}
        \centering
        \includegraphics[width=0.35\textwidth]{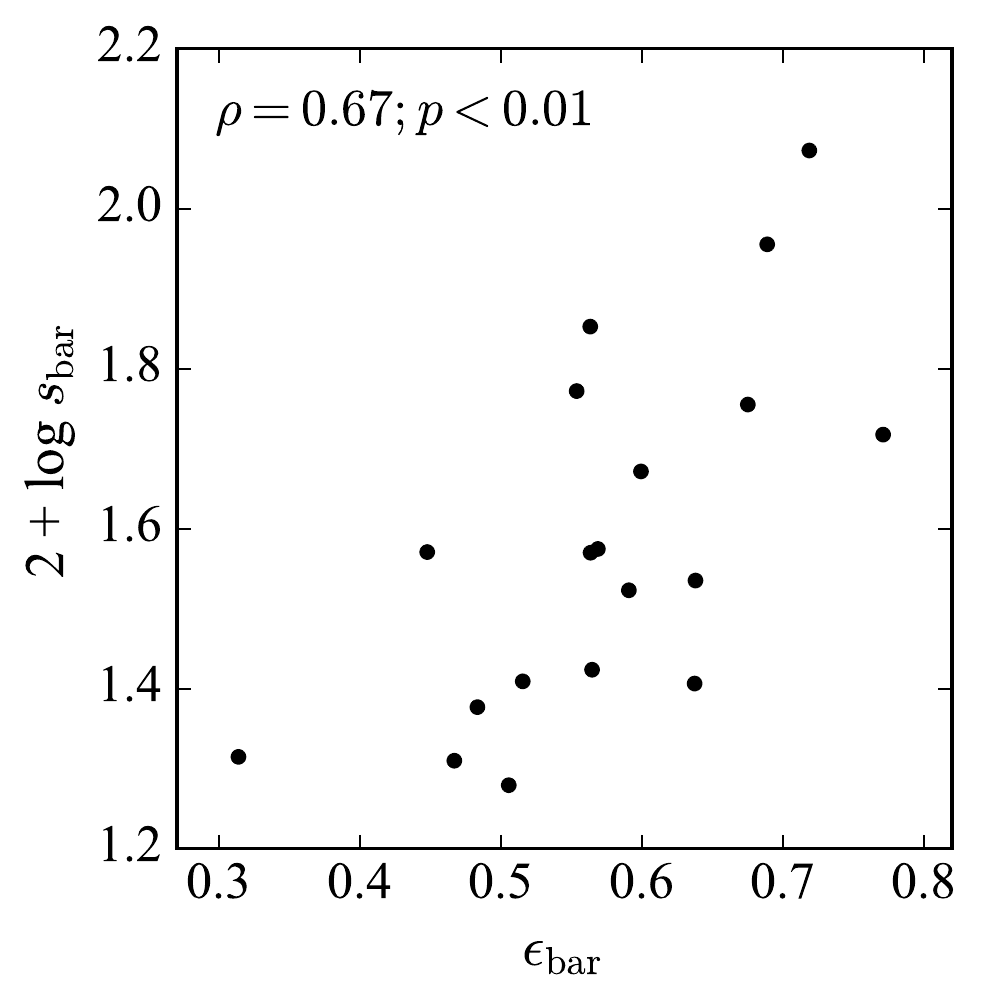}
        \caption{Relation between the bar strength ($2+\log s_{\rm bar}$) and the bar ellipticity ($\epsilon_{\rm bar}$). The Pearson correlation coefficient, $\rho$, and the $p$ value are shown at the top.
        }
        \label{app_ebar}
\end{figure}

\begin{figure}
        \centering
        \includegraphics[width=0.35\textwidth]{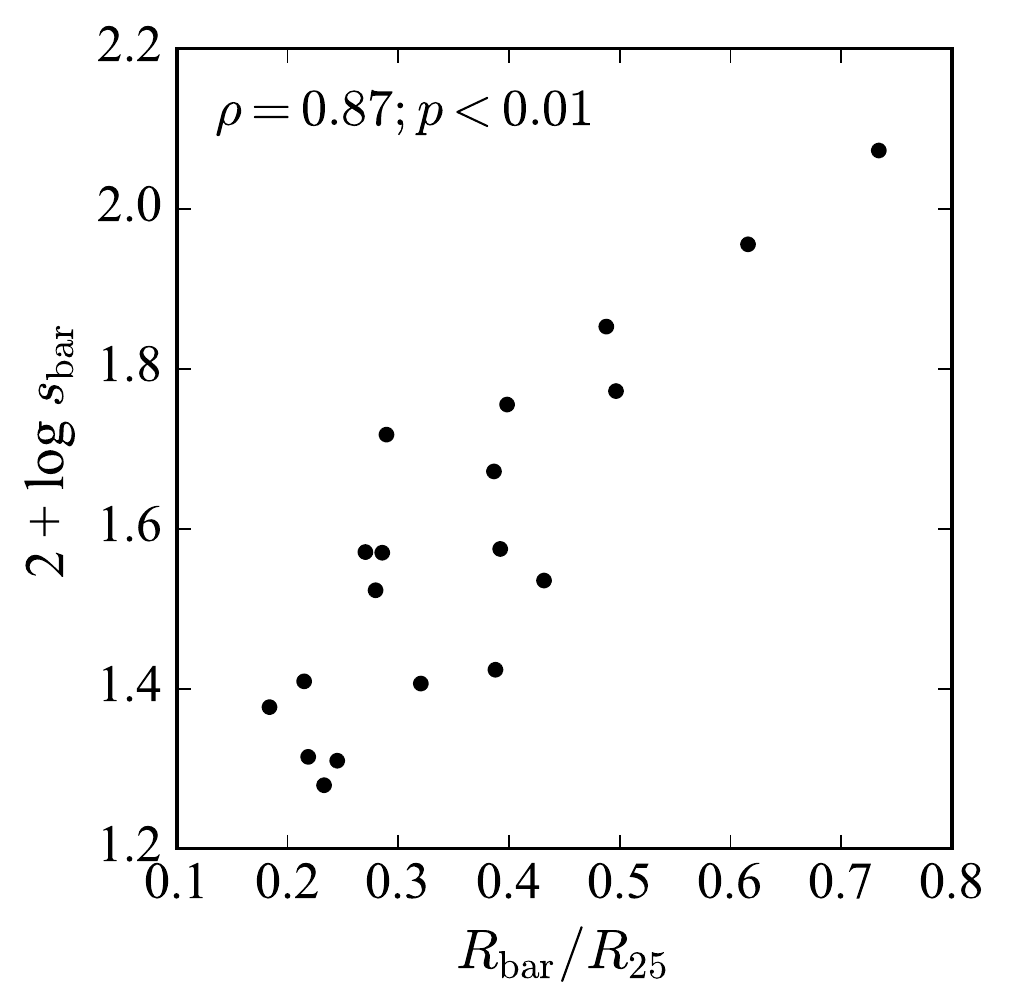}
        \caption{Relation between the bar strength ($2+\log s_{\rm bar}$) and the relative length of the bar ($R_{\rm bar}/R_{25}$). The Pearson correlation coefficient, $\rho$, and the $p$ value are shown at the top.
        }
        \label{app_Lbar}
\end{figure}

The ellipticity and relative length are measures of bar strength \citep{Martin1995, Martinet1997, Aguerri1999, Marinova2007, Barazza2008, Barazza2009, Gadotti2009, Li2011}. It is instructive to compare bar strengths defined in our framework with the measured $\epsilon_{\rm bar}$ and $R_{\rm bar}/R_{25}$. We calculate average relative Fourier amplitude over the region occupied by the bar $s_{\rm bar}$ and then compute the bar strength: $2+\log s_{\rm bar}$. 
Since the $\epsilon_{\rm bar}$ are projected values, we focus on galaxies with inclination angle less than 65$\degr$ to avoid projection effects. As shown in Figs.~\ref{app_ebar} and \ref{app_Lbar}, the $2+\log s_{\rm bar}$ are well correlated with the $\epsilon_{\rm bar}$ and $R_{\rm bar}/R_{25}$. The value of $2+\log s_{\rm bar}$ gets higher with increasing bar ellipticity ($\rho=0.67$ and $p<0.01$) and bar length ($\rho=0.87$ and $p<0.01$). 

%--------------------------------------------------------------------
%--------------------------------------------------------------------

%--------------------------------------------------------------------
%--------------------------------------------------------------------

%--------------------------------------------------------------------
%--------------------------------------------------------------------

\end{appendix}

\end{document}